\documentclass[a4paper,fleqn,usenatbib]{mnras}
\pdfoutput=1
\usepackage{newtxtext,newtxmath}
\usepackage[T1]{fontenc}
\usepackage{ae,aecompl}

\usepackage{graphicx, subfig}	
\usepackage{amsmath}	
\usepackage{amssymb}	

\usepackage{xspace}
\hypersetup{pdfauthor={Trebitsch, Volonteri and Dubois},pdftitle={BH growth obscuration and duty-cycle at high-z}}

\newcommand{\hi}{\ifmmode {\mathrm{H\,\textsc{i}}} \else H\,\textsc{i} \fi}
\newcommand{\hii}{\ifmmode {\mathrm{H\,\textsc{ii}}} \else H\,\textsc{ii} \fi}
\newcommand{\hei}{\ifmmode {\mathrm{He\,\textsc{i}}} \else He\,\textsc{i} \fi}
\newcommand{\heii}{\ifmmode {\mathrm{He\,\textsc{ii}}} \else He\,\textsc{ii} \fi}
\newcommand{\heiii}{\ifmmode {\mathrm{He\,\textsc{iii}}} \else He\,\textsc{iii} \fi}
\newcommand{\Msun}{\ifmmode {\rm M}_{\odot} \else ${\rm M}_\odot$\xspace\fi}
\newcommand{\Mvir}{\ifmmode {M_{\rm vir}} \else $M_{\rm vir}$\xspace\fi}
\newcommand{\Mstar}{\ifmmode {M_\star} \else $M_{\star}$\xspace\fi}
\newcommand{\Mbh}{\ifmmode {M_\bullet} \else $M_{\bullet}$\xspace\fi}
\newcommand{\muv}{\ifmmode {M_{1450}} \else $M_{1500}$\xspace\fi}
\newcommand{\Lbol}{\ifmmode {L_{\rm bol}} \else $L_{\rm bol}$\xspace\fi}
\newcommand{\kms}{\ifmmode {\, \rm km\,s^{-1}} \else $\,{km\,s^{-1}}$\xspace\fi}
\newcommand{\fedd}{\ifmmode {f_{\rm Edd}} \else $f_{\rm Edd}$\xspace\fi}
\newcommand{\NH}{\ifmmode {N_{\rm H}} \else $N_{\rm H}$\xspace\fi}

\newcommand{\ramses}{\textsc{Ramses}\xspace}

\usepackage[usenames,dvipsnames,svgnames,table]{xcolor}

\title[BH obscuration and duty-cycles at high-$z$]{Black hole obscuration and duty cycles mediated by AGN feedback in high redshift galaxies}

\author[M. Trebitsch et al.]{
Maxime Trebitsch$^{1}$\thanks{E-mail: maxime.trebitsch@iap.fr}, 
Marta Volonteri$^{1}$ and
Yohan Dubois$^{1}$
\\
$^{1}$Sorbonne Universit\'{e}, CNRS, UMR 7095, \\
\phantom{$^{1}$}Institut d'Astrophysique de Paris, 98 bis bd Arago, 75014 Paris, France\\
}

\date{Accepted XXX. Received YYY; in original form ZZZ}

\pubyear{2018}
\usepackage{lipsum}
\begin{document}
\label{firstpage}
\pagerange{\pageref{firstpage}--\pageref{lastpage}}
\maketitle

\begin{abstract}
  Dense gas in the centre of galaxies feeds massive black holes, but can also become a source of obscuration and limit our ability to find faint Active Galactic Nuclei (AGN).    
  We use a high resolution cosmological radiative hydrodynamics simulation to connect the properties of the gas in the central region (a few tens of parsecs) of a high redshift galaxy to the growth of a massive black hole during the first billion years of the Universe.
  We find that the feedback from the AGN efficiently controls the growth of the black hole and limits the duration of the high accretion episodes by emptying the gas reservoir. As the galaxy grows in mass, the production of metals results in the presence of dust-enriched gas in the galaxy centre that can obscure highly accreting black holes enough to strongly reduce their UV/optical visibility. We also find that the gas outside the very centre of the galaxy can contribute to the total column density and obscuration at a level at least comparable to the gas in the nuclear region.
  We suggest that this explains the different duty cycles required to explain the masses of high redshift quasars and the observed UV/optical luminosity functions: in our case, the AGN would be observed with an X-ray luminosity above $L_{\rm X} = 10^{42}\,\mbox{erg s}^{-1}$ around 30\% of the time, but with UV magnitude brighter than $M_{1450} = -23$ only 4\% of the time.
  
\end{abstract}




\section{Introduction}
\label{sec:intro}

Active Galactic Nuclei (AGN) are sources emitting over a very large range of wavelengths and, as a family, they display numerous differences. Many of these differences are ascribed to the presence of material between the engine and the observer \citep[][and references therein]{Hickox2018}. The absorbing material absorbs a part of the flux, dimming its brightness and modifying the transmitted spectral energy distribution. Broad emission lines, produced near the black hole (BH), can only be observed if no absorbing material is present, otherwise one observes narrow lines and infrared emission produced by reprocessing of the continuum emitted by the source. 

 The classic unification scheme \citep[][and references therein]{1993ARA&A..31..473A,1995PASP..107..803U,2015ARA&A..53..365N} explains the variety of observed AGN properties with absorbing material distributed non-uniformly, possibly in an axisymmetric or toroidal structure, ``near'' the BH, within the central few parsec. While this picture has had great success in explaining and interpreting most observations, gas in the host galaxy can further modulate the level of obscuration. In particular, high levels of host-wide obscuration are often advocated in conjunction with galaxy mergers \citep{1988ApJ...325...74S,2006ApJS..163....1H}. In the case of powerful quasars the obscured fraction seems to increases with strong far-infrared luminosity, indicating high levels of star formation, while high-luminosity AGN do not seem, necessarily, to be characterized by non-nuclear obscuration \citep{2014MNRAS.437.3550M,2018ApJ...856..154S}. 
 
In local AGN, \citet{Ricci2017} find that the fraction of obscured AGN is almost independent of luminosity, but has a strong dependence on Eddington ratio, with BHs accreting at high rates being less obscured. They advocate for the obscuring gas being close to the BH (within the sphere of influence) and for radiative feedback for the AGN being responsible for clearing out this gas when the BH reaches an effective dust-dependent Eddington rate.   
 
The fraction of obscured AGN appears to increase with redshift \citep{2006ApJ...652L..79T,2014MNRAS.445.3557V}, and this evolution is more evident at high luminosity, $L_{\rm X}\sim 10^{44}-10^{45}\,\mbox{erg.s}^{-1}$ for both optically and X-ray obscured AGN \citep{2014MNRAS.437.3550M,Vito2018}. In \citet{Vito2018} they compared the obscured fraction as a function of luminosity of AGN at $z=3-6$ to the local one and found a strong evolution at high luminosities, with an increase of the fraction from $\sim 10-20\%$ to $\sim 70-80\%$. 
 
High-redshift galaxies are often rich in gas \citep[e.g.][]{Tacconi2010, Tacconi2013, Santini2014}, although not necessarily very dusty or metal rich in their earliest phases of evolution. This gas is not always available for feeding the BH in the nuclear region, because of local supernova feedback \citep{Dubois2015, Bower2017, Angles-Alcazar2017, Habouzit2017, Trebitsch2018} especially in low mass galaxies, but once the galaxy/halo become sufficiently massive and/or compact, the central regions can retain gas and feed the BH. This gas provides the reservoir to feed the accretion disc and then the BH, but it is also a source of obscuration.  

In this paper we examine the role of gas in the near and around the Bondi radius in obscuring radiation from BHs accreting in high-$z$ galaxies. The region we study is intermediate between the ``torus'' (in its various nuances) and the ``nuclear starburst'', beyond the dust sublimation radius and around the gravitational sphere of the BH \citep[see][for a discussion of the various physical scales]{2014MNRAS.437.3550M,Hickox2018}. Previous numerical investigations of obscuration around BHs focused on multi-scale re-simulations of idealized galaxies \citep{2012MNRAS.420..320H,2016MNRAS.458..816H}; the final re-simulations had a higher spatial and mass resolution than our study, but did not include the cosmological evolution of a young galaxy.  We do not fully resolve the broad line region and the putative torus in our simulations, therefore their effects should be added as a correction. In other words, we estimate a lower limit to the total amount of obscuration in young galaxies that are actively star forming without being starbursts but live in very dynamic environments.

\section{Description of the simulation}
\label{sec:sims}

We perform a zoom-in simulation using the public, multigroup radiative transfer (RT) version of the adaptive mesh refinement (AMR) code \ramses \footnote{\url{https://bitbucket.org/rteyssie/ramses/}} \citep{Teyssier2002, Rosdahl2013, Rosdahl2015}.
The collisionless particles (stars and dark matter) are evolved using a particle-mesh method with a cloud-in-cell interpolation. For the gas, \ramses solves the Euler equations with the second-order MUSCL scheme \citep{vanLeer1979} using the HLLC Riemann solver from \citet{Toro1994} and a MinMod total variation diminishing scheme to reconstruct the intercell-fluxes. We impose a Courant factor of 0.8 to define the timestep.

The RT module propagates the radiation emitted by both stars and BHs (modelled as sink particles, see Sect.~\ref{sec:sims:bhagn}) in three frequency intervals describing the \hi, \hei and \heii photon field. The frequencies are defined by the ionization energies of these species, so the groups are 13.60--24.59 eV, 24.59--54.42 eV and we limit the final \heii-ionizing band to the interval 54.42--1000 eV. The radiation is evolved on the AMR grid using a first-order Godunov method to solve the first two moments of the RT equation and assuming the M1 closure \citep{Levermore1984, Dubroca1999} for the Eddington tensor. To maintain the computational cost of the simulation manageable, we utilize the reduced speed of light approximation as implemented in \citet{Rosdahl2013}, and we choose a reduced speed of light of $\tilde{c} = 0.01 c$, which is reasonable provided we do not focus on the propagation of ionization fronts in the IGM.
The coupling to the hydrodynamical evolution of the gas is done through the non-equilibrium thermochemistry for hydrogen and helium, assuming the on-the-spot approximation (any photon emitted after a recombination is absorbed locally). We ignore the radiation pressure from the UV field to the gas, which plays only a subdominant role at galactic scales \citep{Rosdahl2015a}.
Radiation is emitted by star particles (representing fully sampled stellar population) as a function of their age and metallicity following the models of \citet{Bruzual2003}. The AGN emit radiation when the accretion rate onto BH is high. The full description of the model is given in Sect.~\ref{sec:bh:radiation}.

The AMR grid is refined using a quasi-Lagrangian criterion: a cell is selected for refinement if $\rho_{\rm DM} \Delta x^3 + (\Omega_{\rm DM}/\Omega_b)\rho_{\rm gas} \Delta x^3+ (\Omega_{\rm DM}/\Omega_b) \rho_* \Delta x^3 > 8\ m_{\rm DM}^{\rm HR}$, where $\rho_{\rm DM}$, $\rho_{\rm gas}$ and $\rho_*$ are respectively the DM, gas and stellar densities in the cell, $\Omega_{\rm DM}$ and $\Omega_b$ respectively the cosmic DM and baryon mass density, $\Delta x$ is the cell size, and $m_{\rm DM}^{\rm HR}$ is the mass of the highest resolution DM particle. In a DM-only run, this criterion would allow refinement as soon as there are at least 8 high-resolution DM particles in a cell.

\subsection{Initial conditions}
\label{sec:sims:ics}

In this work, we zoom on a halo that ends up with a mass of $\sim 3 \times 10^{11} \Msun$ at redshift $z \sim 5.7$ embedded in a cosmological volume of $40 h^{-1}$ comoving Mpc on a side. The halo undergoes a $\sim$ 4:1 merger starting around $z \sim 7.1$ which translates into a galaxy merger around $z \sim 6.3$. The halo has been selected from a DM-only run of the full cosmological volume, initially run with $512^3$ particles ($m_{\rm DM}^{\rm LR} \simeq 5 \times 10^7\, \Msun$). The initial conditions for the cosmological box as well as the zoom region have been created with \textsc{Music}\footnote{\url{https://bitbucket.org/ohahn/music/}} \citep{Hahn2011} for a flat $\Lambda$CDM cosmology consistent with the \emph{Planck} results \citep[dark energy density $\Omega_\Lambda = 0.692$, total matter density $\Omega_m = 0.308$, Hubble parameter $h = 0.6781$ and baryon matter density $\Omega_b = 0.048$,][]{Planck2015}. We select the target halo in the final output with \textsc{HaloMaker} \citep{Tweed2009}, which uses the \textsc{AdaptaHOP} algorithm \citep{Aubert2004}.

The initial conditions for the halo of interest are then regenerated at an effective resolution of $4096^3$ elements (level $\ell = 12$), and the external region is degraded by one level so that the box is covered by a grid with $256^3$ elements (level $\ell = 8$). This translates in a mass resolution of $m_{\rm DM}^{\rm LR} \simeq 4 \times 10^8\, \Msun$ for the lowest resolution dark matter particles, and $m_{\rm DM}^{\rm HR} \simeq 10^5\, \Msun$ for the high-resolution particles.
For the full radiative hydrodynamics run, we then allow for refinement down to a minimum cell size of $\Delta x = 40 h^{-1}\mbox{Mpc}/2^{23} \simeq 7\,\mbox{pc}$. The gas in the simulation is assumed to be neutral and homogeneously metal poor, with an initial gas phase metallicity $Z = 5\times 10^{-3} Z_\odot = 10^{-4}$.

\subsection{Star formation and feedback}
\label{sec:sims:sffb}
At the resolution of our simulation, we describe the stars forming in the galaxies as star particles with mass $m_\star \sim 1.8\times 10^4\,\Msun$.

\subsubsection{Star formation}
\label{sec:sims:star-formation}
We model star formation with a Schmidt-like law \citep{Schmidt1959}, with an approach similar to that \citet{Rasera2006}. This thermo-turbulent model has been previously described in details in \citet{Kimm2017, Trebitsch2017}, and we use the same version as implemented in \citet{Trebitsch2018}.

During one timestep $\Delta t$, we convert a gas mass $M_{\rm sf} = \epsilon \rho_{\rm gas} \Delta x^3 \Delta t/t_{\rm ff}$ in star-forming cells, where $G$ is the gravitational constant, $\epsilon$ is the local star formation efficiency and $t_{\rm ff} = \sqrt{3\pi / 32 G \rho_{\rm gas}}$ is the free fall time of the gas. The exact number $N$ of star particles formed in $\Delta t$ is drawn from a Poisson distribution $P(N) = (\lambda^N/N!) \exp(-\lambda)$ of parameter $\lambda = M_{\rm sf} / m_\star$.

Cells are considered star forming when the local density $\rho_0$ exceeds a threshold $\rho_{\rm threshold} = 1\,\mbox{cm}^{-3}$, chosen as the typical ISM density, and when the local turbulent Mach number exceeds $\mathcal{M} \geq 2$.
In each of these cells (assumed to correspond to a star forming cloud), the efficiency $\epsilon$ is computed following the `multi-ff PN' model of \citet{Federrath2012}, based on \citet{Padoan2011}.
\begin{equation}
  \label{eq:sfr_ff}
  \epsilon \propto \text{e}^{\frac{3}{8}\sigma_s^2}\left[1 + \mathrm{erf}\left(\frac{\sigma_s^2 - s_{\rm crit}}{\sqrt{2\sigma_s^2}}\right)\right],
\end{equation}
where $\sigma_s  = \sigma_s(\sigma_{\rm gas}, c_{\rm s})$ characterizes the turbulent density fluctuations, $s_{\rm crit} \equiv \mathrm{ln}\left(\frac{\rho_{\rm gas, crit}}{\rho_0}\right)$ is the critical density above which the gas will be accreted onto stars, and $\rho_{\rm gas, crit} \propto (\sigma_{\rm gas}^2 + c_{\rm s}^2) \frac{\sigma_{\rm gas}^2}{c_{\rm s}^2}$.

\subsubsection{Feedback from massive stars}
\label{sec:sims:feedback}
Our simulation models feedback from massive stars through two main channels: type II supernovae (SNe) and radiative feedback resulting from photoionization heating.
Supernovae explosions are modelled using the subgrid implementation first detailed in \citet{Kimm2014} and \citet{Kimm2015}.
We deposit mass and momentum in the cells neighbouring the SN host cell in a single event $t_{\rm SN} = 5\,\mbox{Myr}$ after each star particle is formed. The amount of momentum released depends on the local density and metallicity of each neighbouring cell in order to capture correctly the momentum transfer at all stages of the Sedov blast wave.
Photoionization heating is directly included through the coupling between the ionizing photon field and the gas.
Following \citet{Kimm2017}, when the Str\"omgren sphere of a star particle is not resolved, we increase the final radial momentum from SNe due to the pre-processing of the ISM by radiation before the SN explosion, as suggested by \citet{Geen2015}.

\subsection{BH model}
\label{sec:sims:bhagn}

Just as in \citet{Trebitsch2018}, the BH seeding as well as the subsequent growth and associated feedback are modelled following the fiducial implementation of \citet{Dubois2012}. We will now briefly recall the main features of this model, and describe two new features used in this work for the BH dynamics and for the radiation produced by the AGN.

\subsubsection{BH seeding}
\label{sec:bh:seeding}
We represent SMBH using sink particles with initial mass $M_{\bullet,0} = 3 \times 10^4\,\Msun$. These sink particles are created in cells where the following criteria are met: both the gas and stellar density exceeds a threshold that we choose to be $\rho_{\rm sink} = 100\,\mbox{cm}^{-3}$; the gas must be Jeans-unstable, and there should be enough gas in the cell to form the sink particle. We also impose an exclusion radius $r_{\rm excl} = 40\,\mbox{kpc}$ to avoid the formation of multiple SMBH in the same galaxy. Each sink particle is then dressed with a swarm of `cloud' particles equally spaced by $\Delta x/2$ on a regular grid lattice within a sphere of radius $4 \Delta x$. These cloud particles provide a convenient way of probing and averaging the properties of the gas around the BH.

\subsubsection{Accretion}
\label{sec:bh:accretion}
Once a sink particle is formed, it is allowed to grow via gas accretion and BH-BH mergers. Mergers are allowed to happen when two BH are closer than $4\Delta x$ from one another if their relative velocity is lower than the escape velocity of the binary system they would form.
Gas accretion is implemented following the classical Bondi-Hoyle-Lyttleton formula \citep{Bondi1952}
\begin{equation}
\dot{M}_{\rm BHL} = 4\pi G^2 \Mbh^2 \frac{\bar{\rho}}{\left(\bar{c}_s^2 + \bar{v}_{\rm rel}^2\right)^{3/2}}
\label{eq:accrate-bhl}
\end{equation}
where $\Mbh$ is the BH mass, $\bar{\rho}$ and $\bar{c}_s$ are respectively the average gas density and sound speed, and $\bar{v}_{\rm rel}$ the relative velocity between the BH and the surrounding gas. The bar notation denotes an averaging over the cloud particles.
The quantities are averaged using a Gaussian kernel $w \propto \exp\left(-r^2/r_{\rm sink}^2\right)$, where $r_{\rm sink}$ is defined using the Bondi radius $r_{\rm BHL} = \frac{G\Mbh}{\bar{c_s}^2 + \bar{v}_{\rm rel}^2}$:
\begin{equation}
  \label{eq:rsink}
  r_{\rm sink} = 
  \begin{cases}
    \Delta x/4 & \text{if}\quad r_{\rm BHL} < \Delta x/4, \\
    r_{\rm BHL}  & \text{if}\quad \Delta x/4 \leq r_{\rm BHL} < 2 \Delta x, \\
    2 \Delta x & \text{if}\quad 2 \Delta x \leq r_{\rm BHL}.
  \end{cases}
\end{equation}
Just as in \citet{Trebitsch2018}, we do not use any artificial boost for the gas accretion onto the BH. The accretion rate is capped at the Eddington rate:
\begin{equation}
\dot{M}_{\rm Edd} = \frac{4\pi G \Mbh m_p}{\epsilon_r \sigma_{\rm T} c},
\label{eq:accrate-edd}
\end{equation}
where $m_p$ is the proton mass, $\sigma_{\rm T}$ is the Thompson cross section, $c$ is the speed of light and $\epsilon_r$ is the radiative efficiency of the accretion flow onto the BH $\epsilon_r = 0.1$. The final BH accretion rate is:
\begin{equation}
  \label{eq:accrate-both}
  \dot{\Mbh} = \min\left(\dot{M}_{\rm BHL}, \dot{M}_{\rm Edd}\right),
\end{equation}
and we further prevent the BH to accrete more than 25\% of the gas content of the cell in one timestep for numerical stability.

\subsubsection{Dynamics}
\label{sec:bh:dynamics}
As discussed e.g. in \citet{Pfister2017}, dynamical friction plays a large role in setting the detailed dynamics of a BH moving in a galaxy. We however do not have the resolution in our simulation to resolve properly this phenomenon. 
We include a drag force to model the unresolved dynamical friction exerted by the gas lagging behind the BH as introduced in~\citet{Dubois2013}. This will mostly play a role at high redshift, when the galaxies are gas rich and clumpy, and will help to maintain the BH in dense gas clumps without resorting to regularly re-positioning the BH. The frictional force is proportional to $F_{\rm DF} = \alpha f_{\rm gas} 4\pi \rho (G \Mbh / \bar{c_s}^2)$, where $\alpha$ is an artificial boost, with $\alpha  = (\rho/\rho_{\rm DF, th})^2$ if $\rho > \rho_{\rm th}$ and 1 otherwise, and $f_{\rm gas}$ is a fudge factor varying between 0 and 2 which depends on the BH Mach number $\mathcal{M}_\bullet = \bar{v}_{\rm rel}/\bar{c}_s$ \citep{Ostriker1999, Chapon2013}. In this work, we take $\rho_{\rm DF, th} = 50\,\mbox{cm}^{-3}$. Additionally, we include a second contribution to the dynamical friction, this time caused by the collisionless particles (stars and DM). The implementation, based on the analysis of \citet{Chandrasekhar1943, Binney1987} will be described in Pfister et al. (in prep), and is somewhat similar to that of \citet{Tremmel2015}.

\begin{figure*} 
  \centering
  \includegraphics[width=.75\linewidth]{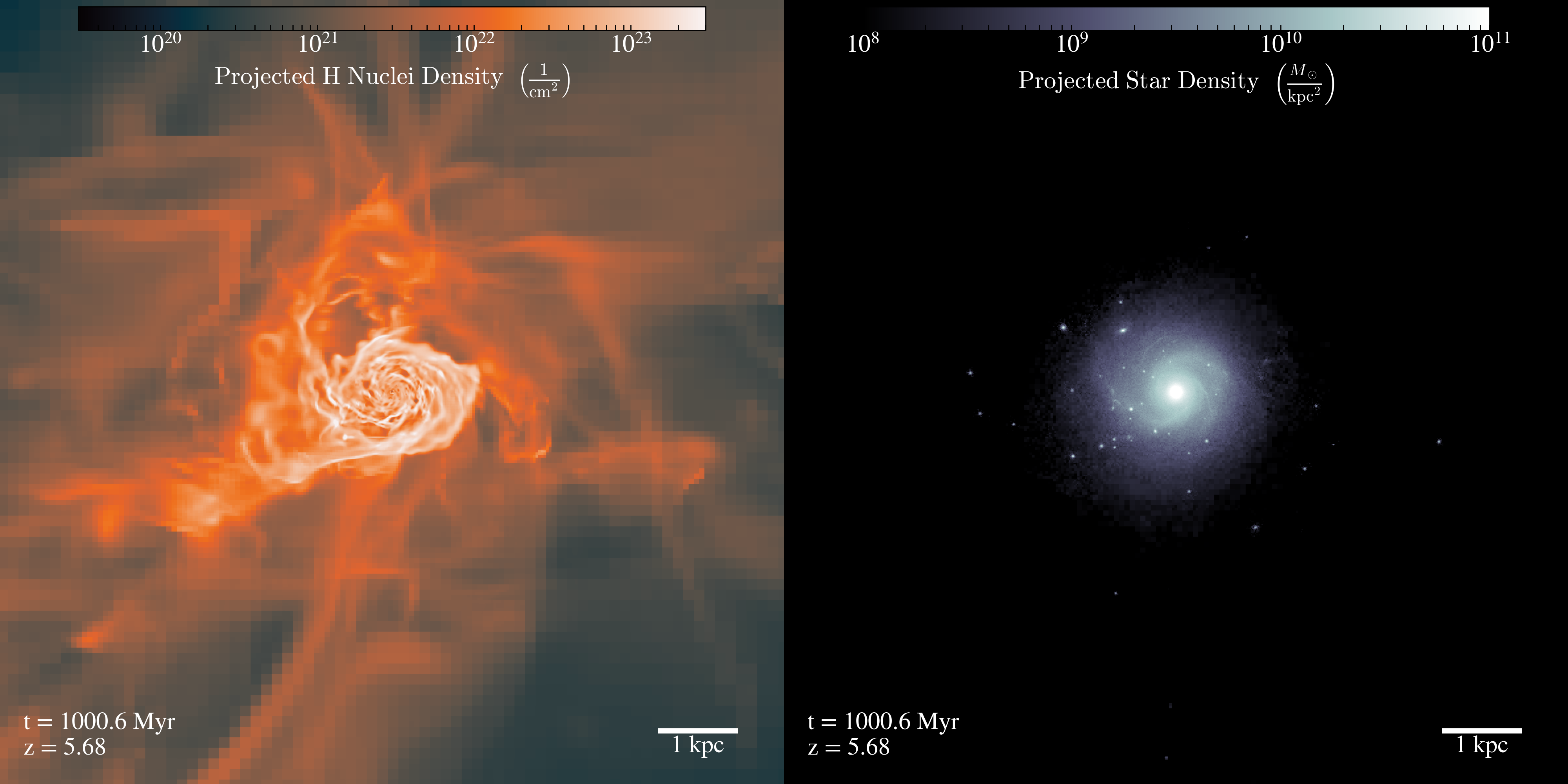}
  \caption{Illustration of the main galaxy studied in this work at the end of the simulation, $z \simeq 5.7$, with the gas column density on the left and the stellar surface density on the right.}
  \label{fig:illustration}
\end{figure*}

\subsubsection{AGN feedback}
\label{sec:bh:feedback}

AGN feedback resulting from the accretion onto a BH is modelled using the dual mode implementation of \citet{Dubois2012}. At low Eddington ratio $\lambda_{\rm Edd} = \dot{M}_{\rm BHL} / \dot{M}_{\rm Edd} < 0.01$, the AGN is in ``radio mode'', and in ``quasar mode'' when $\lambda_{\rm Edd} \geq 0.01$.
For the quasar mode, we release thermal energy in a sphere of radius $\Delta x$ centred on the BH at a rate $\dot{E}_{\rm AGN} = \epsilon_f \epsilon_r \dot{M}_\bullet c^2$, where $\epsilon_f = 0.15$ is the fraction of the radiated energy that is transferred to the gas, and $\dot{\Mbh}$ is the BH accretion rate.
For the radio mode, we deposit momentum as a bipolar outflow with a jet velocity $u_{\rm J} = 10^4\ \mbox{km\,s}^{-1}$, modelled as a cylinder of radius $\Delta x$ and height $2\Delta x$ weighted by a kernel function
\begin{equation}
  \label{eq:jetprofile}
  \psi\left(r_{\rm cyl}\right) = \frac{1}{2\pi\Delta x^2}\exp\left(-\frac{r_{\rm cyl}^2}{\Delta x^2}\right),
\end{equation}
with $r_{\rm cyl}$ the cylindrical radius. Mass is removed from the central cell and deposited in the cells enclosed by the jet at a rate $\dot{M}_{\rm J}$ with
\begin{equation}
  \label{eq:jetmass}
  \dot{M}_{\rm J}(r_{\rm cyl}) = \frac{\psi(r_{\rm cyl})}{\Psi} \eta_{\rm J} \dot{M}_\bullet
\end{equation}
where $\Psi$ is the integral of $\psi$ over the cylinder and $\eta_{\rm J} = 100$ is the mass-loading factor of the jet accounting for the interaction between the jet and the ISM at unresolved scales. Gas is pushed away from the BH on each side of the cylinder with the norm of the momentum $q(r_{\rm cyl})$ given by
\begin{equation}
  \label{eq:jetmomentum}
  q_{\rm J}(r_{\rm cyl}) = \dot{M}_{\rm J}(r_{\rm cyl}) u_{\rm J},
\end{equation}
and we inject the corresponding kinetic energy into the gas:
\begin{equation}
  \label{eq:jetegy}
  \dot{E}_{\rm J}(r_{\rm cyl}) = \frac{q_{\rm J}(r_{\rm cyl})^2}{2 \dot{M}_{\rm J}(r_{\rm cyl})} = \frac{\psi(r_{\rm cyl})}{\Psi} \eta_{\rm J} \dot{E}_{\rm AGN}.
\end{equation}
The jet is aligned with the total angular momentum of the accreted gas, and the radio mode efficiency is assumed to be $\epsilon_f = 1$. The feedback efficiencies in both the radio and quasar modes are empirically determined in order to reproduce the BH-to-bulge mass relations at $z=0$ \citep{Dubois2012}.

\subsubsection{AGN radiation}
\label{sec:bh:radiation}
Finally, we have developed a radiative feedback implementation for the AGN when it is in quasar mode, similar to the work of \citet{Bieri2017} and to the Appendix~C of \citet{Trebitsch2018}. We release radiation at each fine timestep, and the amount of radiation released in each frequency bin is given by the luminosity of the quasar in each band. Contrary to \citet{Bieri2017} and \citet{Trebitsch2018}, we do not use the averaged \citet{Sazonov2004} spectrum for the BH but instead use a piecewise power-law  corresponding to a \citet{Shakura1973} thin disc extended by a power-law at high energy. In each band, the AGN luminosity is therefore a function of $\Mbh$ and $\dot{\Mbh}$ (formally, there is also a spin dependence, but we do not follow the spin of the BH in this work). We further assume that $f_{\rm IR} = 30\%$ of the bolometric luminosity of the disc is absorbed by dust and re-emitted as IR radiation, that we do not model here, and we keep the radiative efficiency of the accretion flow to $\epsilon_r = 0.1$.

We ensured that the adopted AGN SED yielded an average spectrum similar to the AGN SED used in \citet{Volonteri2017}. However, in the remainder of this work, we only use this SED to estimate the ionizing luminosity of the AGN.

\subsection{Gas cooling and heating}
\label{sec:sims:cooling}

Our simulation includes non-equilibrium cooling for hydrogen and helium by tracking the abundances of H, H$^+$, He, He$^+$, He$^{++}$, complemented by metal cooling modelled using the standard rates tabulated in \ramses and computed with \textsc{Cloudy}\footnote{\url{http://www.nublado.org/}} {\citep[last described in][]{Ferland2017}} above $10^4$~K. Below $10^4$~K, we account for energy losses via metal line cooling following \citet{Rosen1995}, where the effect of the metallicity is included by scaling linearly the metal cooling enhancement. We do not take into account the impact of the local ionizing flux on metal cooling, but instead assume photo-ionization equilibrium with a redshift dependent \citet{Haardt1996} UV background for the metals. This UV background is not used for the hydrogen and helium non equilibrium photo-chemistry, which is done using the local photon field transported self-consistently by the RT solver.

\section{Results}
\label{sec:results}

\subsection{Global properties}
\label{sec:global-properties}

\begin{figure}
  \centering
  \includegraphics[width=\linewidth]{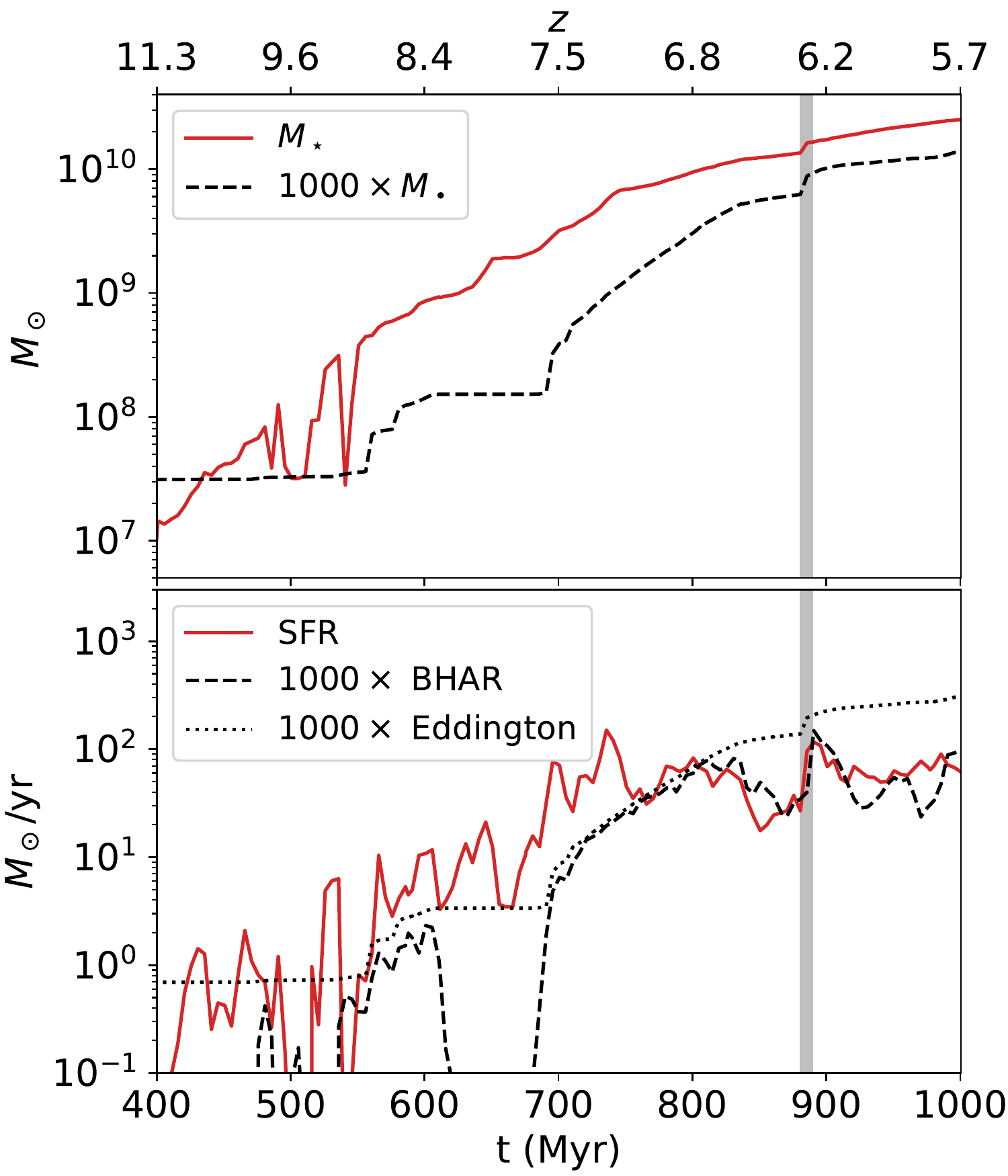}
  \caption{\emph{Top}: BH mass and stellar mass as a function of time for the most massive BH and galaxy of the simulation. \emph{Bottom}: SFR and BH accretion rate (smoothed over $\sim 5$ Myr), as well as the Eddington accretion rate as a function of time. The grey-shaded area marks a major merger. The BH-galaxy matching is done purely based on the distance of the BH to the closest galaxy, resulting in some artefacts at early times.}
  \label{fig:mstar-mbh}
\end{figure}

We run \textsc{AdaptaHOP} on the star particles to identify galaxies in the simulation. The main galaxy is identified at the last timestep ($z \sim 5.7$, illustrated on Fig.~\ref{fig:illustration}), and we track back its progenitors using \textsc{TreeMaker} \citep{Tweed2009}. The central BH at that time is then matched to the its host galaxy at each timestep. We show in Fig.~\ref{fig:mstar-mbh} the evolution of both the host galaxy (red solid line) and the BH (black dashed line), with the upper panel displaying the evolution of the mass of the galaxy and the BH, and the lower panel showing the star formation rate (SFR) and the BH accretion rate (BHAR). The vertical grey line indicates the last major merger for the host galaxy.

When the galaxy is settled and well identified ($z \lesssim 9$), the SFR slowly increases from $\dot{M}_\star \simeq 10\,\Msun \mbox{yr}^{-1}$ to $\dot{M}_\star \simeq 80\,\Msun \mbox{yr}^{-1}$ for a stellar mass growing from $10^{9}\Msun$ to almost $M_\star \simeq 2.5 \times 10^{10}\,\Msun$. The specific SFR oscillates around $1 - 5\,\mbox{Gyr}^{-1}$, and strongly peaks at the time of the major merger ($z \sim 6.3$) as the SFR reaches $\dot{M}_\star \gtrsim 200\,\Msun \mbox{yr}^{-1}$.
The dip in the BH accretion rate at $t \simeq 620 - 680\,\mbox{Myr}$ is due to the BH being ejected from the central region of the galaxy due to a minor merger.
This also explains the stagnation of the BH mass around $\Mbh \simeq 1.5\times 10^5\,\Msun$ during that time. When the BH falls back to the galaxy, it merges with another BH of similar mass to reach $\Mbh \simeq 3\times 10^5\,\Msun$ and then steadily grows until the end of the simulation, where it reaches $\Mbh \simeq 1.4\times 10^7\,\Msun$.

\subsection{Self-regulation of the local column density}
\label{sec:NH-accretion-feedback}

Let us now focus on the properties of the gas surrounding the BH in the central region of the galaxy in order to examine the accretion-feedback cycle at play when the BH is growing.

In the simulation, accretion is estimated through the Bondi formalism, therefore we can rewrite Eq.~\ref{eq:accrate-bhl} as a relation between accretion rate, or Eddington ratio, and column density:
\begin{equation}
\dot{\Mbh}=\frac{4 \pi G^2 m_p \Mbh^2 \NH}{R_{\rm acc}\left(\bar{c}_s^2 + \bar{v}_{\rm rel}^2\right)^{3/2}},
\label{eq:NHacc_bondi}
\end{equation}
where $\NH$ is the local gas column density around the BH measured in the accretion kernel of radius $R_{\rm acc} = 4\Delta x$. From this, we can write the dimensionless Eddington ratio $\fedd = \dot{\Mbh}/\dot{M}_{\rm Edd}$ as
\begin{equation}
\fedd= 7.58 \, \epsilon_{r,0.1} \Mbh_{,6} \langle N_{\rm H,22} \rangle R_{\rm acc,40}^{-1}(c_{\rm s,10}^2 + v_{\rm rel,10}^2)^{-3/2},
\label{eq:NHacc}
\end{equation}
where $\epsilon_{r,0.1}$ is the radiative efficiency normalized to 0.1,  $\Mbh_{,6}$ is the BH mass in units of $10^6\,\Msun$, $\langle N_{\rm H,22}\rangle$ is the average gas column density in units of $10^{22} \, \mbox{cm}^{-2}$ within the accretion radius, $R_{\rm acc,40}$ is the radius within which the accretion rate is estimated, in units of $40~{\rm pc}$, and $c_{\rm s,10}$ and $v_{\rm rel,10}$ are the sound speed and relative velocity between gas and BH in units of $10 \kms$. Note that in the simulation \fedd is capped at unity.

From Eq.~\ref{eq:NHacc}, it appears that a BH needs to be surrounded by dense gas (i.e., by a large average column density) in its vicinity to be able to accrete at high rate. If this were the only physics at play, one should expect a linear correlation between \fedd and $\langle N_{\rm H}\rangle$  at fixed BH mass, temperature and dynamical state. Since temperature and density are not independent, with dense gas being colder than rarefied gas, and relative velocity between the gas and the BH are also a function of the dynamical state, the relation we measure in the simulation is close to $N_{\rm H,22} \propto \fedd^{0.50}$. In this analysis, we exclude BHs at $\fedd=1$ to avoid being biased by the cap. Including the BHs at $\fedd=1$ only marginally affects the trend, and yields $N_{\rm H,22} \propto \fedd^{0.56}$. In any case, a higher \fedd corresponds to a higher $\langle\NH\rangle$ and vice-versa.

We now move to the effect of feedback, through which an accreting BH injects energy in its surroundings. In our simulation, three types of feedback are included: mechanical energy injection at low \fedd, thermal energy injection at high \fedd, and radiation from the AGN at high \fedd.
Since we do not include Compton heating from X-rays in the simulation, we do not expect the radiation to play a huge role at the scales we resolve here. Indeed, the photoionization heating from the AGN will at most increase the local gas temperature to a few $10^4$ K, and would maintain the local sound speed above 10 km/s. In the simulation, the effective sound-speed $v_{\rm Bondi} = \sqrt{\bar{c}_s^2 + \bar{v}_{\rm rel}^2}$ is most of the time above 10 km/s.
We therefore limit ourselves to the case of direct energy injection, where we can estimate the relation with gas column density by taking the ratio of the gas binding energy to the injected energy:
\begin{align}
  \frac{E_b}{E_{\rm inj}} & \sim \frac{G (\Mbh + M_\star + M_{\rm DM} + M_{\rm gas}) M_{\rm gas}}{R_{\rm acc}} \frac{1}{\epsilon_f \epsilon_r \Delta t \fedd L_{\rm Edd}} \nonumber\\
                          & \sim 1.9 \times 10^{-4} \frac{\langle N_{\rm H,22}\rangle}{\fedd} \frac{M_{\rm tot}}{\Mbh} \frac{R_{\rm acc,40}}{\epsilon_{r,0.1}\epsilon_{f,0.15} \Delta t_{0.1}}
\label{eq:energy}
\end{align}
with $M_{\rm gas}=4 \pi m_p \langle \NH\rangle R_{\rm acc}^2/3$. Here $\Delta t\sim 0.1$~Myr is the coarse timestep in the simulation (hence $\Delta t_{0.1}$ is normalized to 0.1~Myr) and $\epsilon_{f,0.15}$ is the feedback efficiency normalized by 15\%, the efficiency of the thermal energy injection. In this derivation, we have defined $M_{\rm tot}$ as the total enclosed mass within $R_{\rm acc}$, but in practice, we do not resolve the influence radius of the BH, and at the resolution of our simulation the gravitational potential is dominated by the stellar component in the vicinity of the BH.
This quantity can be seen as a stability criterion for the central gas concentration: when $E_b/E_{\rm inj} < 1$, the AGN is powerful enough to evacuate the gas from the central region and stopping the accretion onto the BH. At fixed Eddington ratio, Eq.~\ref{eq:energy} shows that the higher the column density the AGN has to lift, the less effective the feedback will be at unbinding the central gas concentration.
We should also keep in mind that dense gas can cool rapidly and thus some of the injected energy is ineffective in unbinding the gas.  Recall that for radiative cooling, dominant on the scales and densities of relevance here, $t_{\rm cool} \propto \rho^{-1}\propto \NH^{-1}$. We do not include the correction here, but the injected energy should be rescaled by the fraction of gas that can cool in one timestep. We also note that Eq.~\ref{eq:energy} is formally only valid at $\fedd > 0.01$, because the jet feedback model at lower accretion rate is, by definition, not isotropic.

\begin{figure}
  \centering
  \includegraphics[width=\linewidth]{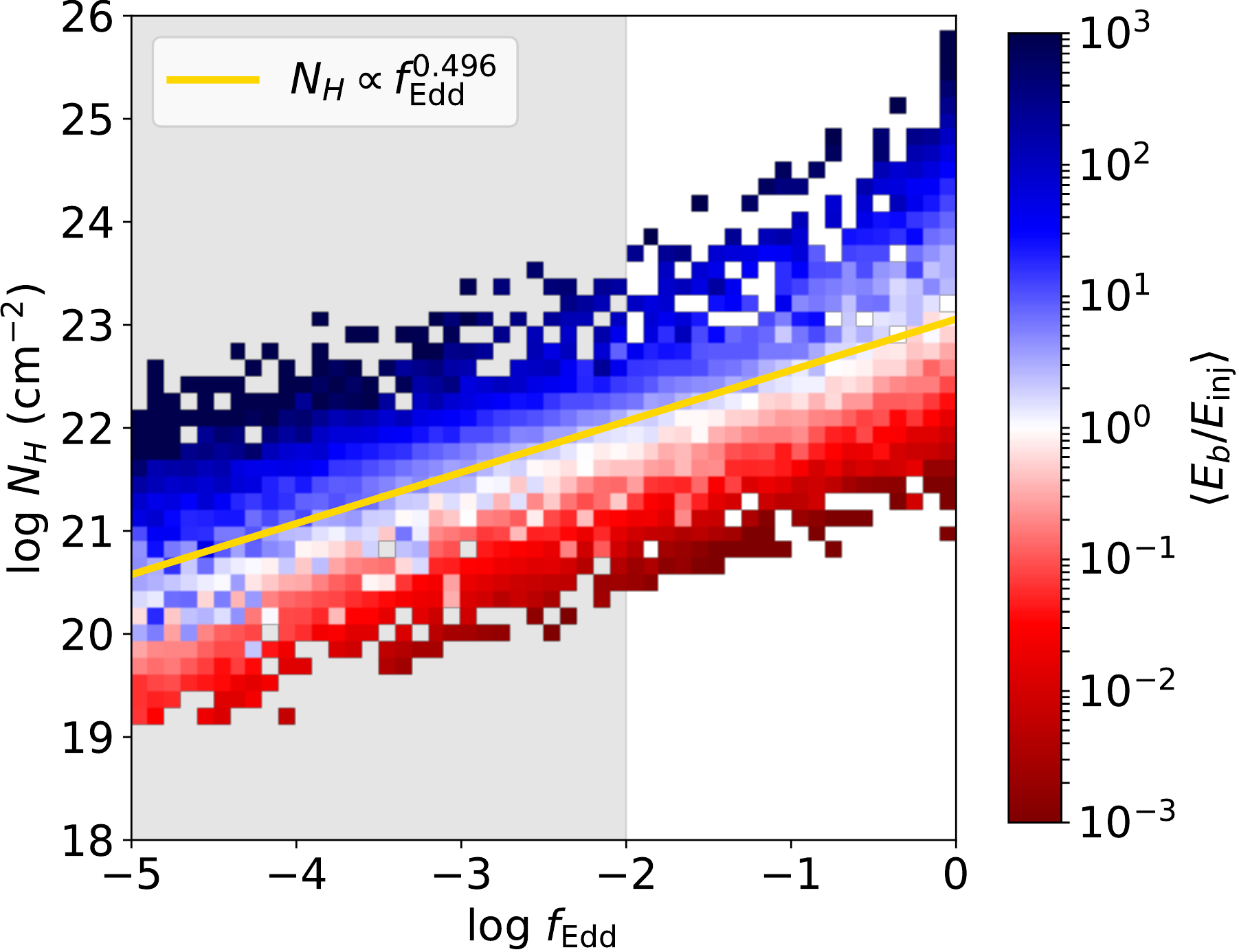}
  \caption{Distribution of Eddington ratio and column density in a sphere with radius 40~pc, colour-coded by the average ratio of binding energy of the gas in that sphere and the injected energy from AGN feedback. When $E_b/E_{\rm inj}$ is larger than unity, the AGN is unable to unbind the gas supply that should continue feeding it. The grey area marks $\fedd < 10^{-2}$, where the AGN is jet mode and this analysis is not strictly valid. The yellow line indicates the power-law fit of \NH as a function of \fedd.}
  \label{fig:NH-fedd_EbEinj}
\end{figure}

In Fig.~\ref{fig:NH-fedd_EbEinj}, we show the two-dimensional distribution\footnote{Note that in this figure as well as all the following (unless mentioned explicitly), we stack all the snapshots together. This effectively smooths any trend that would depend on redshift, but these trend would not be well captured by a single zoom simulation.} of the column density \NH against the Eddington ratio \fedd in a sphere of 40~pc around the BH, colour-coded by the ratio of the binding energy of the gas to the injected energy from AGN feedback (assuming that the feedback mode is always thermal for simplicity; the grey shaded region marks where this assumption formally breaks down).
The yellow line shows the $N_{\rm H,22} \propto \fedd^{0.496} \sim \fedd^{0.5}$ fit discussed previously. In the red (blue) area, the feedback from the AGN is (not) strong enough to remove the gas from the centre.
The general behaviour is perhaps unsurprising: at fixed column density, the BH is unable to unbind the gas when \fedd is low, and conversely at fixed \fedd, the higher the column density and the easier it is for the gas to resist the feedback.
It is interesting to note that at fixed \fedd, the average column density (indicated by the yellow line) is just slightly higher than that for which $E_b = E_{\rm inj}$ (the white region). This can be understood in terms of AGN feedback self-regulating the BH growth: if the BH is able to remove all of the gas that it feeds on ($E_b < E_{\rm inj}$), the accretion rate will drop, and as a result the injected energy from AGN feedback will drop as well. From Eq.~\ref{eq:NHacc}, we see that AGN feedback reduces $\fedd$ faster than $N_{\rm H}$. Indeed, the AGN-driven wind will not only remove some gas, it will also heat whatever remains, therefore reducing further $\fedd$ For example, after a feedback event, if \NH decreases by a factor 10, \fedd will decrease by more than a factor 10, because the sound speed and relative velocity will increase as well. If the BH falls in a state where $E_b > E_{\rm inj}$, then the surrounding gas will be able to cool and feed the BH again. This will be followed by a new episode of feedback. As a result of this feeding-feedback cycle, $E_b / E_{\rm inj}$ will hover around unity most of the time, naturally leading to an average behaviour of $N_{\rm H,22} \propto \fedd^{0.5}$. During the last phase of the simulation ($t \gtrsim 800$ Myr), when the BH self-regulate its growth, we have a series of episodes where $E_b \lesssim E_{\rm inj}$ followed by short episodes where the BH stops accreting, resulting in $E_{\rm inj} \rightarrow 0$ and therefore $E_b/E_{\rm inj} \rightarrow \infty$.

This overall paints a picture in which AGN feedback is effectively regulating the gas content in the vicinity of the central BH: when there is too much gas in the centre, the BH will accrete efficiently and release enough energy through feedback to reduce the density and slow down the BH growth. This translates into a modulation of the column density around the BH, that we discuss in the following.

\subsection{Obscuration and local column density}
\label{sec:obsc-column-dens}

\begin{figure}
  \centering
  \includegraphics[width=\linewidth]{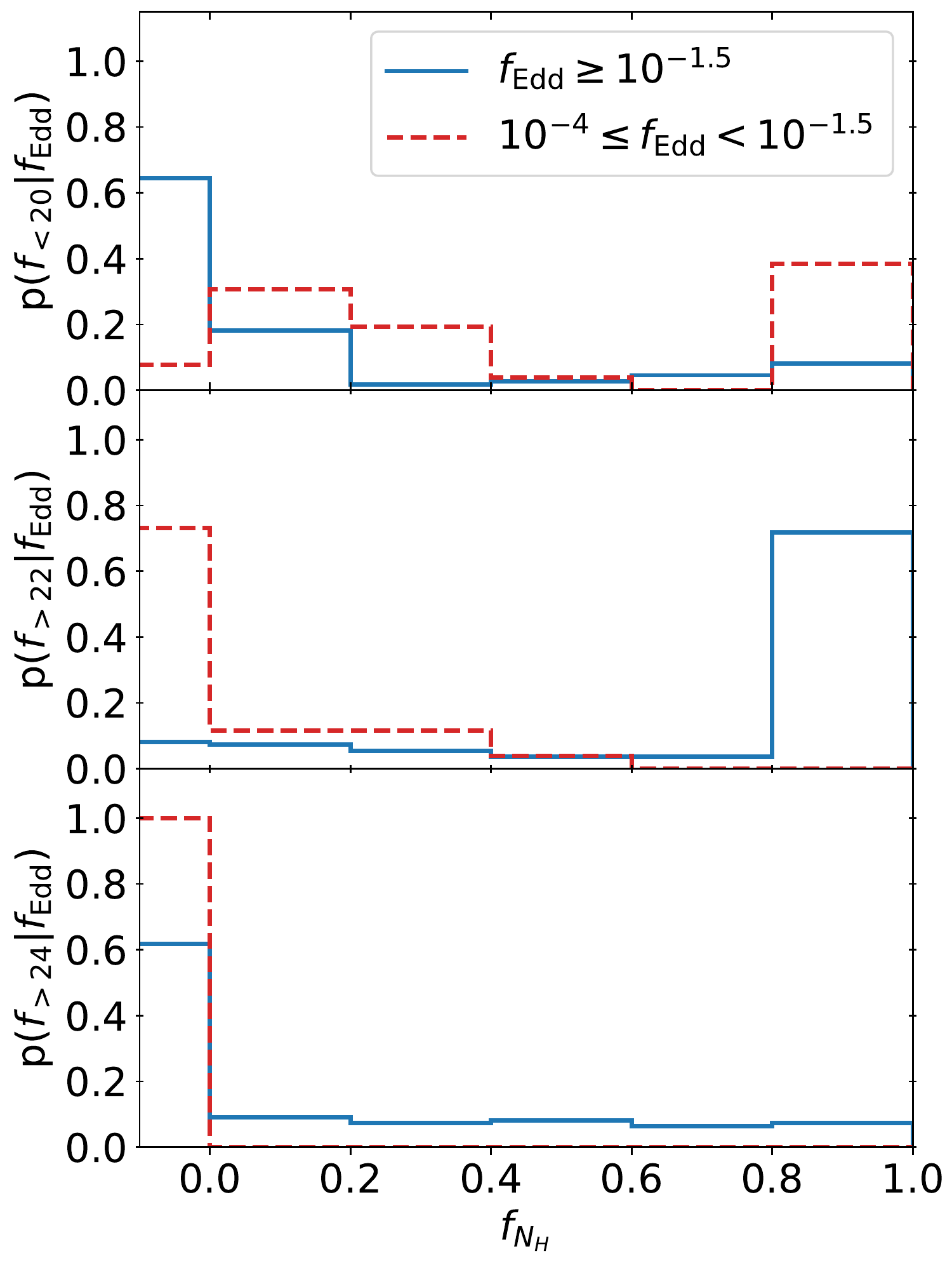}
  \caption{Distributions of the fraction of lines of sight above or below given column density thresholds, in bins of Eddington ratio, i.e. at low accretion rates ($10^{-4}<\fedd<10^{-1.5}$) and high accretion rates ($\fedd>10^{-1.5}$). \emph{Top}: fraction with $\NH<10^{20}\, \mbox{cm}^{-2}$. \emph{Middle}: fraction with $\NH>10^{22}\, \mbox{cm}^{-2}$. \emph{Bottom}: fraction with $\NH>10^{24}\, \mbox{cm}^{-2}$. A fraction shown below zero represents the fraction at exactly zero. Low accretors ($10^{-4}<\fedd<10^{-1.5}$) have zero lines of sight with $\NH>10^{24}\, \mbox{cm}^{-2}$. Conversely, high accretors with $\fedd>10^{-1.5}$ have a 65\% probability of not having fully transparent ($\NH<10^{20}\, \mbox{cm}^{-2}$) lines of sights.}
  \label{fig:fracNH-distr}
\end{figure}

We will now discuss the spatial distribution of obscuring material around the BH. To this end, we measure the column density of neutral gas around the BH integrating the local density along rays starting from the BH position and with a length 40 pc. We use a total of 192 rays, and the direction of each ray is drawn from a \textsc{HealPix}\footnote{\url{https://healpix.sourceforge.io/}} \citep{Gorski2005} tessellation of the sphere.

\begin{figure}
  \centering
  \includegraphics[width=\linewidth]{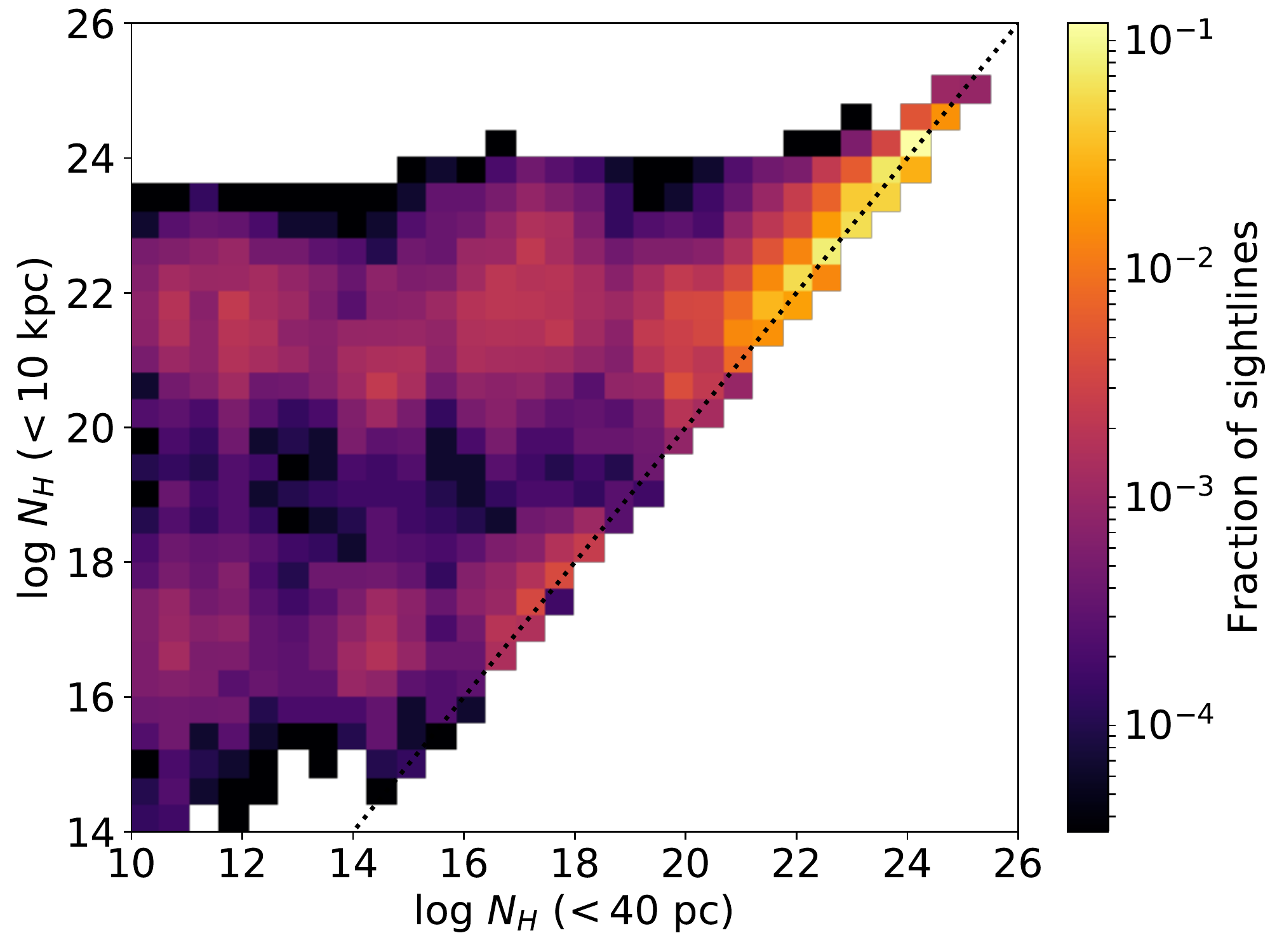}
  \caption{Comparison between the column density around the BH and for the whole galaxy, for all the snapshots of the simulation. Even when the AGN is unobscured, there will still be absorbing gas the rest of the galaxy. In total, around 14\% of the sightlines have $\NH_{,\rm gal} > 2\, \NH_{,\rm loc}$, and more than 1\% have $\NH_{,\rm gal} > 8\, \NH_{,\rm loc}$.}
  \label{fig:NHloc_NHgal}
\end{figure}

\begin{figure*}
  \centering
  \includegraphics[width=.9\linewidth]{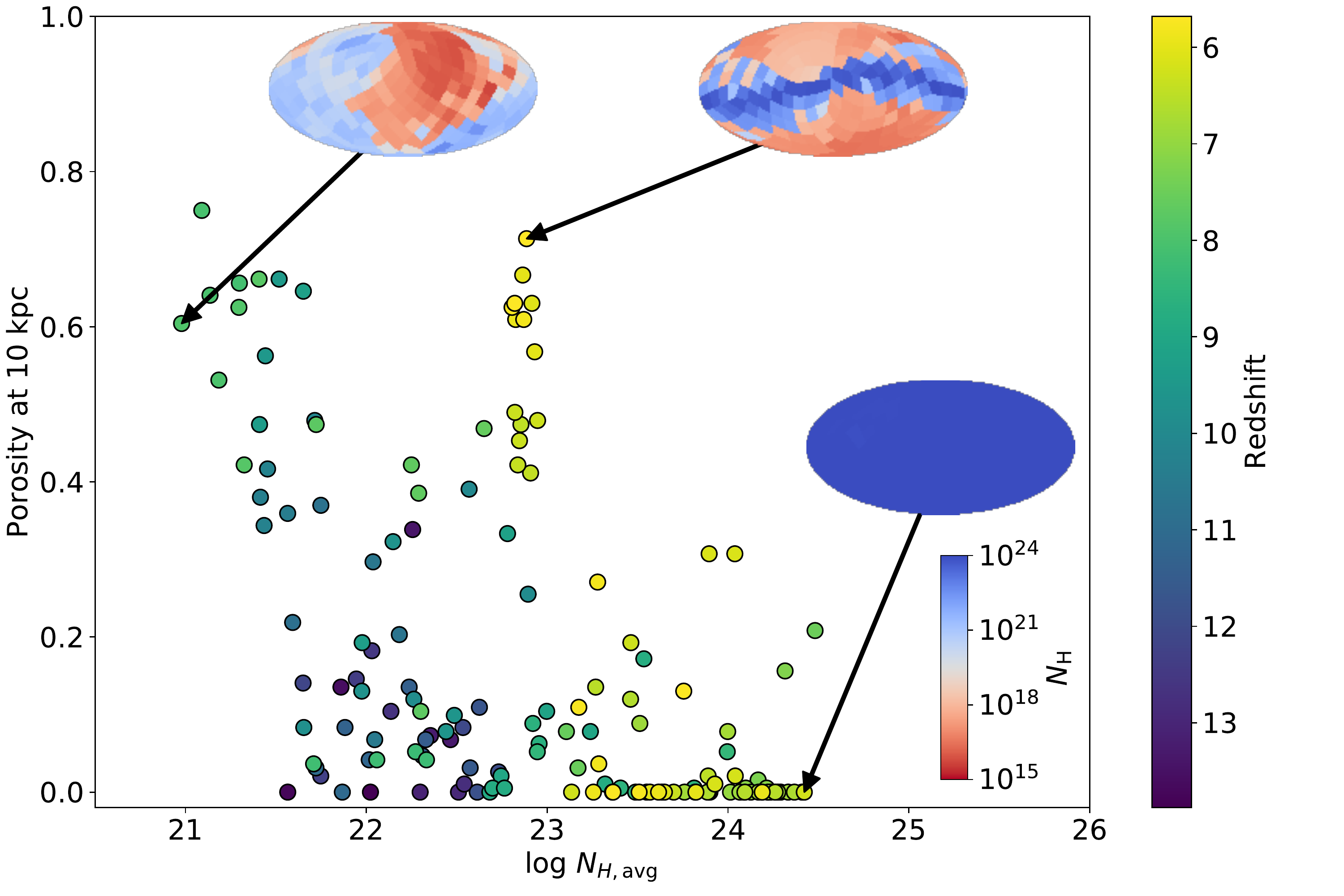}
  \caption{Porosity (defined as the fraction of lines of sight with \NH below a fifth of the average $\NH_{\rm avg}$) as a function of $\NH_{\rm avg}$. The three insets represent all-sky projections of \NH around the BH, with a colour scale increasing from red to blue.}
  \label{fig:porosity}
\end{figure*}
We show in Fig.~\ref{fig:fracNH-distr} the distribution of the fraction of sightlines from the BH, above or below given column densities ($f_{>24}$ is the fraction above $10^{24} \mbox{cm}^{-2}$, $f_{>22}$ is the fraction above $10^{22} \mbox{cm}^{-2}$, $f_{<20}$ is the fraction below $10^{20}\, \mbox{cm}^{-2}$) when the central BH is accreting efficiently (in blue) or not (in red). At higher accretion rates, the BH is more often surrounded by a large fraction of dense gas. This is especially striking in the middle panel of Fig.~\ref{fig:fracNH-distr}: at high accretion rate, the BH is almost always surrounded by a column density larger than $\NH \geq 10^{22}\, \mbox{cm}^{-2}$, and this is almost never the case for BH accreting inefficiently.
This is in contrast with \citet{Ricci2017}, who analyse the \emph{Swift}/BAT sample of local AGN and find an anti-correlation between column density and Eddington ratio. \citet{Ricci2017} interpret their results as obscuration being caused by gas within a few tens of parsecs from the BH, with radiative feedback from dusty gas in this inner region responsible for clearing our gas around the most accretion BHs \citep[see also][]{1991MNRAS.252..586L}. They suggest that BHs will loop through a series of high and low accretion rates events, where the feedback from the AGN would clear out the gas obscuring the BH (reducing \NH) while it is still accreting efficiently from a gas reservoir, therefore leading to a large fraction of unobscured AGN at high Eddington ratio. Once the gas reservoir is emptied, \fedd decreases and the AGN will become obscured again as the gas falls back to the BH vicinity.
While the feedback process they describe happens on scales smaller than our resolution, we confirm, numerically and analytically, that a BH moves to higher \NH as its Eddington ratio increases and that feedback modulates the column density of gas around the BH. In our simulation, however, accretion on the BH is determined instantaneously from the local gas density, without an additional accretion reservoir, therefore we cannot directly assess the time over which a BH can continue accreting after feedback has cleared its surroundings of gas.

Because our simulation includes the galactic and circum-galactic context, we can reproduce a similar analysis for the obscuring material not only in the inner region of the galaxy, but all the way to the circum-galactic medium. For this, we cast again 192 rays from the central BH of the galaxy at each snapshot of the simulation, and integrate the neutral column density over 10 kpc. Fig.~\ref{fig:NHloc_NHgal} compares the column density measured in inner 40 pc around the BH ($\NH_{,\rm loc}$) and in a 10 kpc sphere ($\NH_{,\rm gal}$). The dotted lines marks the $\NH_{,\rm loc} = \NH_{,\rm gal}$ relation. We can see that for a large part, the galaxy-scale column density (and so the obscuring material) is completely dominated by the small-scale \NH, especially in the $\NH > 10^{23}\,\mbox{cm}^{-2}$ regime. However, at lower central \NH (and notably below $10^{20}\,\mbox{cm}^{-2}$), there is a significant fraction of the sightlines where the obscuration comes from the gas beyond the centre of the galaxy.
More quantitatively, around 14\% of all the sightlines have $\xi = \frac{\NH_{,\rm gal}}{\NH_{,\rm loc}} > 2$, and more than 1\% have $\NH_{,\rm gal} > 8\, \NH_{,\rm loc}$. If we further select only the sightlines with $\NH_{,\rm loc}$ above (below) $10^{22}\,\mbox{cm}^{-2}$, the fraction of sightlines where $\xi > 2$ becomes $\sim 8\%$ ($\sim 40\%$).
Outside of the central region, the gas distribution is far from homogeneous (e.g. when the galactic disc forms, as show in Fig.~\ref{fig:illustration}). This explains the scatter in $\NH_{,\rm loc}$ at fixed $\NH_{,\rm gal}$: an AGN for which we would measure a (galaxy-scale integrated) \NH of order $10^{22}\,\mbox{cm}^{-2}$ could either correspond to a BH embedded in a dense enough clump (and would therefore accrete efficiently from a high $\NH_{,\rm loc}$ reservoir) or to a BH that has cleared its local environment through AGN feedback (low $\NH_{,\rm loc}$).
In the first case, we would expect that the BH is surrounded by dense gas in all directions, while in the second case, the observed \NH should depend on the line of sight.
We quantify this by defining the porosity of the medium around the BH, defined by the fraction of sightlines with $\NH < \NH_{,\rm avg}/5$, where $\NH_{,\rm avg}$ is the average column density in the 10 kpc sphere around the BH.
We illustrate the dependence of the porosity as a function of $\NH_{,\rm avg}$ in Fig.~\ref{fig:porosity}, and the colour-coding indicate the redshift of each snapshot. The three insets correspond to the all-sky projections of the column density around the BH at three specific timesteps of the simulation, with \NH increasing from red to blue. The projection is oriented so that the angular momentum of the gas accreted onto the BH is aligned vertically.
The figure present three distinct regions: at very high $\NH_{,\rm avg}$, the porosity is almost always low, meaning that the BH is embedded in a fairly homogeneous clump, as we expected.
The low end of the $\NH_{,\rm avg}$ distribution corresponds to $z \geq 8$, when the galactic disc is not settled at all, and this even includes the event during which the BH spends some time out of the galaxy. The porosity is mostly high, meaning that the BH will be surrounded by a turbulent distribution of gas with a mixture of holes and clumps, as suggested by the inset.
The third region is defined by the points at relatively high $\NH_{,\rm avg}$, but also high porosity: this matches what we described earlier, where the AGN has cleared the local obscuring material, but in certain lines of sight, the galactic disc (the yellow horizontal structure in the inset) still contributes to the total \NH.

\subsection{AGN luminosity}
\label{sec:agn-luminosity}

\begin{figure}
  \centering
  \includegraphics[width=\linewidth]{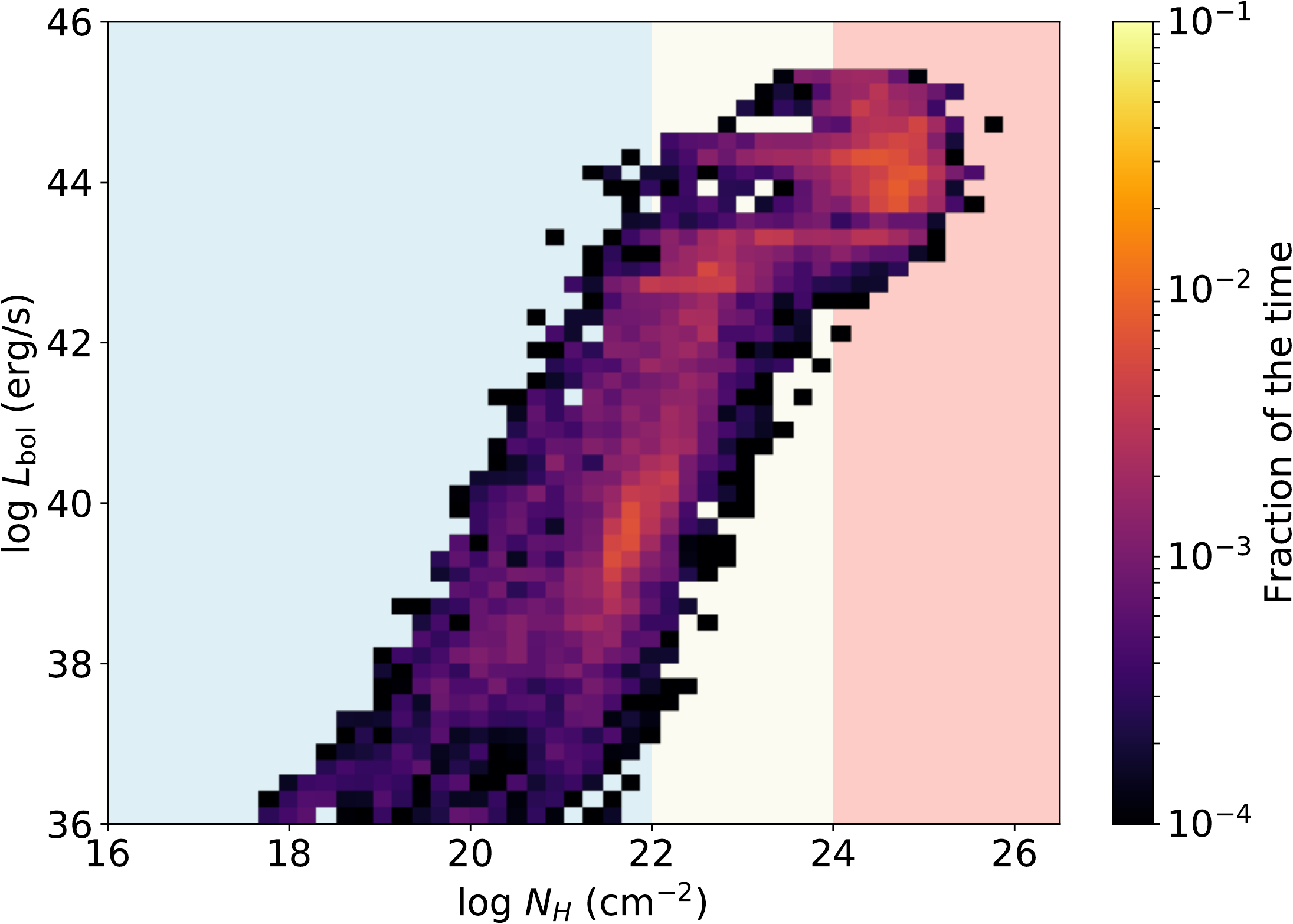}
  \caption{Bolometric luminosity ($\Lbol = 0.1 \dot{M}_\bullet c^2$) as a function local column density in a 40~pc sphere around the BH. The background colours indicate three regime of obscuration considered in this work: obscured in yellow ($\NH > 10^{22}\,\mbox{cm}^{-2}$), Compton-thick in red ($\NH > 10^{24}\,\mbox{cm}^{-2}$), and blue for the rest.}
  \label{fig:Lbol-NH}
\end{figure}

From the discussion in Sect.~\ref{sec:NH-accretion-feedback}, we expect that the accretion rate onto the BH globally increases with the local column density around the BH. This also holds for the AGN bolometric luminosity, defined as $\Lbol = \epsilon_r \dot{M}_\bullet c^2$.
Fig.~\ref{fig:Lbol-NH} shows the bolometric luminosity in the simulation as a function of the local column density in a sphere of 40 pc (the accretion region). The background colours mark three regimes of obscuration: blue for an unobscured AGN ($\NH < 10^{22}\,\mbox{cm}^{-2}$), yellow for an obscured AGN ($\NH < 10^{24}\,\mbox{cm}^{-2}$), and red for Compton-thick AGN ($\NH > 10^{24}\,\mbox{cm}^{-2}$).
While we do not run any X-ray photon package to estimate directly the X-ray luminosity from the simulation, we use the \citet{Hopkins2007} bolometric correction to convert the bolometric luminosity measured in the simulation to the X-ray luminosity in the $[2-10]$ keV band, through $\Lbol = k L_{\rm X}$, with
\begin{equation}
  \label{eq:hopkins_kx}
  k(\Lbol) = 10.83 \left(\frac{\Lbol}{10^{10} L_\odot}\right)^{0.28} + 6.08 \left(\frac{\Lbol}{10^{10} L_\odot}\right)^{-0.020}
\end{equation}
so that $L_{\rm X} = \Lbol/k$.

\begin{figure}
  \centering
  \includegraphics[width=\linewidth]{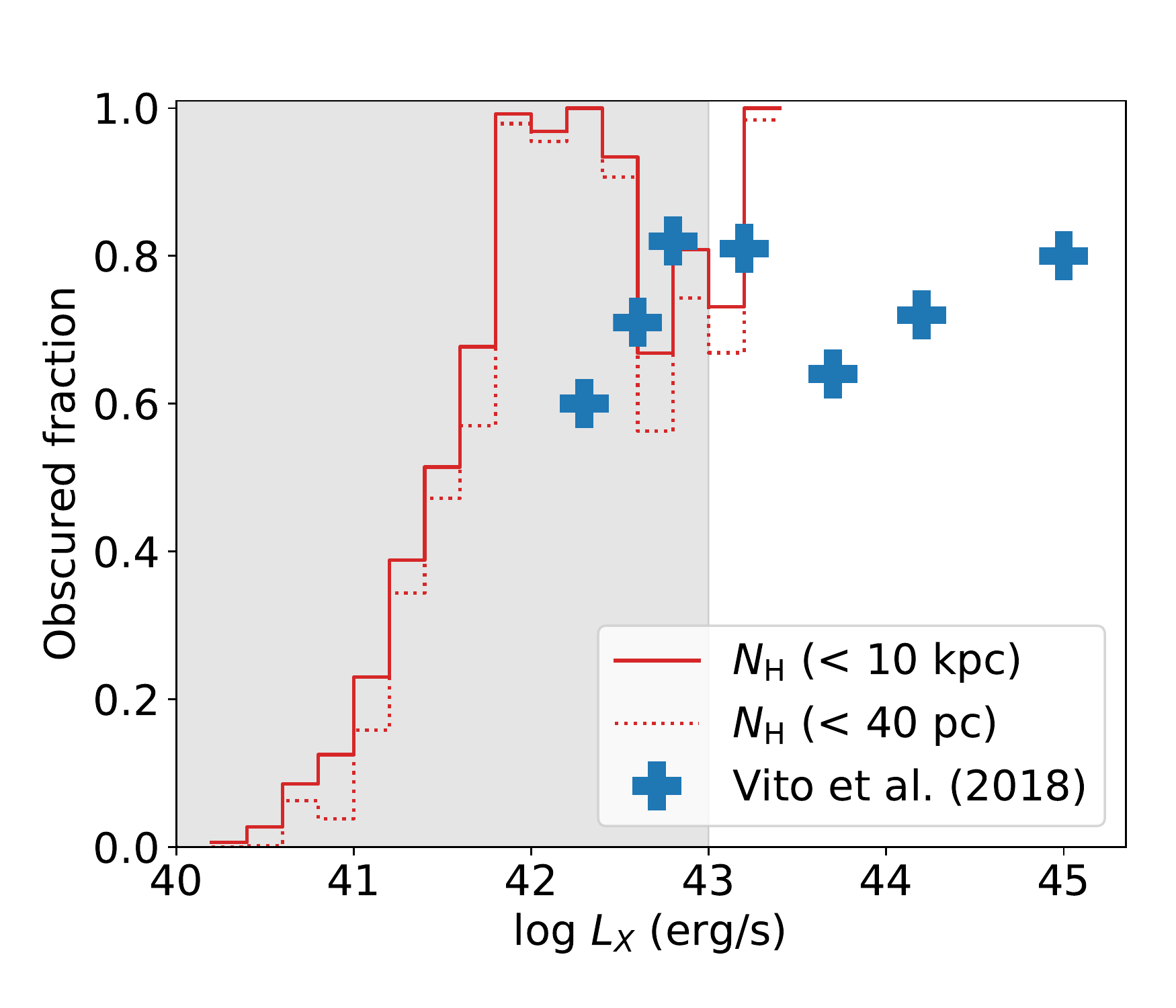}
  \caption{AGN obscured fraction (with column density above $10^{23}\, \rm cm^{-2}$) measured in the simulation (red histogram) compared to the high redshift observations of \citet{Vito2018}, accounting for obscuring material out to 10 kpc (solid line) and 40 pc (dotted line) . We find a trend consistent with high-$z$ observations, with a higher obscured fraction at higher luminosity, and an overall high obscured fraction.}
  \label{fig:fobsc_Vito18}
\end{figure}

With these definitions, we can quantify the AGN obscured fraction in our simulation, defined here as the fraction of AGN lines of sight with column densities above $10^{23}\, \mbox{cm}^{-2}$, following the work of \citet{Vito2018}. Fig.~\ref{fig:fobsc_Vito18} shows the obscured fraction as a function of the intrinsic X-ray luminosity using the column density measured within 10 kpc from the BH. On both panels, the grey shaded region indicates the incompleteness limit from \citet{Vito2018}.
Overall, we find that the obscured fraction increases with luminosity, in qualitative agreement with \citet{Vito2018}, but in contradiction with the study of e.g. \citet{2011ApJ...728...58B} who find the opposite.
\citet{Vito2018} studies a sample of X-ray selected AGN at $z=3-6$ in a parameter space similar to that of the simulation: high-$z$ BHs down to low luminosities. They find, in agreement with our results, a high obscured fraction at high luminosity. Below $10^{43}\, \mbox{erg\,s}^{-1}$ their sample is incomplete, but after correcting for incompleteness, the obscured fraction appears to decrease at low-luminosity, as in our simulation. They note that at low redshift the obscured fraction anti-correlates with luminosity, as found by \citet{Ricci2017} and \citet{2011ApJ...728...58B}. \citet{Vito2018} note also that the obscured fraction overall increases with redshift, and suggest that the differences with respect to the low-redshift Universe are caused by the larger gas fractions of galaxies at earlier cosmic epochs. Our results support this interpretation.
For this simulation, we find that while the galaxy-wide gas contributes somewhat to the total column density of obscuring material (see Fig.~\ref{fig:NHloc_NHgal}), the obscuration is often dominated by the nuclear gas. In particular, the results of Fig.~\ref{fig:fobsc_Vito18} are essentially unchanged if we consider only the obscuring material within 40 pc from the BH (shown as dotted lines). This is in part due to the fact that in this simulation, the AGN only reaches luminosities comparable to that of the sample of \citet{Vito2018} when the BH is in a regime where the two column densities (locally and on a galactic scale) are comparable or dominated by the nuclear gas. Recently, Circosta et al. (submitted), reported observations of obscured AGN at $z > 2.5$ where the ISM of the galaxy is dense enough to contribute significantly to the obscuration of the central AGN. While our results cannot be extrapolated to higher luminosity, our findings are qualitatively consistent with their study: for most of their objects, they find that the ISM column density is at least comparable to the nuclear column density.
We would also expect that the ISM contribution to the obscuration decreases at low redshift, as in the local Universe, only (U)LIRGs reach gas surface density comparable to high-$z$ normal galaxies \citep[e.g.][]{Daddi2010}. This would explain the dichotomy between the results of \citet{Vito2018} and e.g. \citet{2011ApJ...728...58B}: the large gas content of high-$z$ massive galaxies would both significantly contribute to the AGN obscuration and easily replenish the nuclear reservoir once it has been emptied by AGN feedback. This is exactly what our simulation suggest, although at lower AGN luminosity.

We can use a similar approach to convert the AGN bolometric luminosity to a rest-frame UV magnitude at 1450 \AA. Namely, we invert the recommended quasar bolometric correction of \citet{Runnoe2012}:
\begin{equation}
  \label{eq:runnoe}
  \log\left(1450L_{1450}\right) = \frac{\log\Lbol - 4.74}{0.91}.
\end{equation}
So far, we only considered the intrinsic UV or X-ray luminosity of the AGN: however, we have just show in Figs.~\ref{fig:Lbol-NH} and \ref{fig:fobsc_Vito18} that the BH can be significantly obscured. For the X-rays, we can directly compute the obscuration via $\tau_X = \sigma_{\rm T} \NH$. In the UV, the attenuation mainly comes from dust absorption. We estimate the corresponding wavelength-dependent optical depth, $\tau_d(\lambda)$, by integrating the dust column density along the rays used to compute \NH (see Sect.~\ref{sec:obsc-column-dens}):
\begin{equation}
  \label{eq:taudust}
  \tau_d(\lambda) = \int_{\rm ray} n_d(\ell) \sigma_d(\lambda) d\ell,
\end{equation}
with $\sigma_d$ the dust interaction cross section and $n_d$ the number density of dust grains. We follow the implementation of the \textsc{Rascas} code (Michel-Dansac et al., in prep) based on the approach of \citet{Laursen2009} to estimate these quantities: $\sigma_d$ is defined as the cross section per hydrogen atom, using the fits of \citet{Gnedin2008} for the wavelength dependence based on their Small Magellanic Cloud (SMC) model. The dust number density is then to be understood as a pseudo number density, given by $n_d = (n_{\hi} + f_{\rm ion} n_{\hii})Z/Z_0$, where $f_{\rm ion} \sim 0.01$ accounts for the presence of dust in ionized gas, and $Z_0 = 0.005$ the mean metallicity of the SMC. We then evaluate the dust optical depth at 1450 \AA\ as $\tau_{1450} = \tau_d(1450\,\mbox{\AA})$.
\begin{figure}
  \centering
  \includegraphics[width=\linewidth]{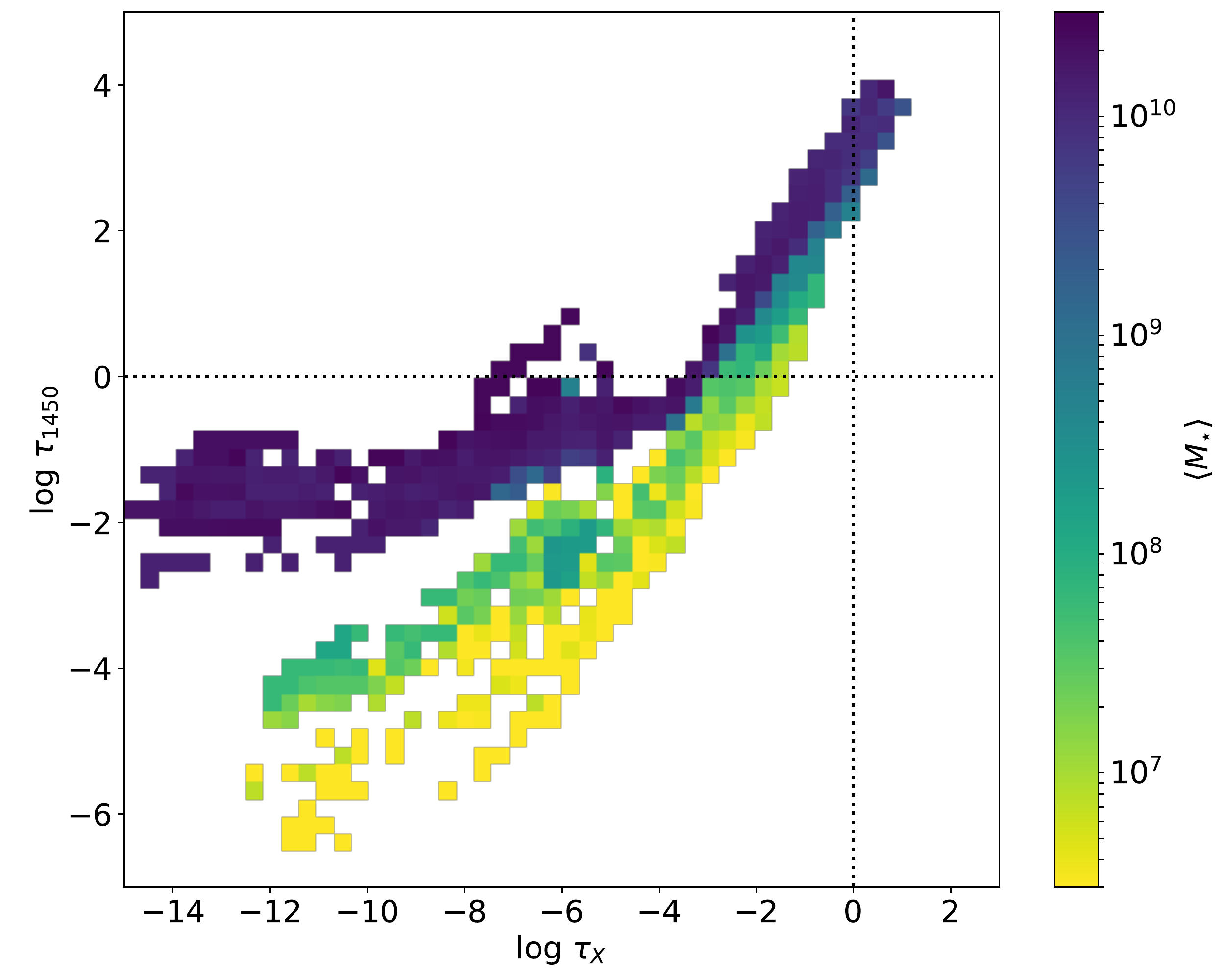}
  \caption{Comparison of the optical depth in the UV ($\tau_{1450}$) due to dust grains with the optical depth in the X-ray band ($\tau_X$); the colour coding indicate the average stellar mass of the galaxy in each bin. A higher stellar mass corresponds to a more evolved (and more metal rich) galaxy, resulting in a higher dust attenuation at fixed \NH.}
  \label{fig:tauX_taudust}
\end{figure}
Fig.~\ref{fig:tauX_taudust} shows the distribution of $\tau_X$ against $\tau_{1450}$, colour-coded by the stellar mass of the galaxy at each timestep. The dotted line indicate $\tau = 1$, which corresponds to an attenuation of roughly a factor three ($\exp(-\tau) \sim 0.37$). As the colour goes from yellow to purple, the galaxy grows in mass and the gas around the BH will be more and more metal-rich: this explains why at fixed $\tau_X$, the dust optical depth $\tau_{1450}$ increases.

\begin{figure}
  \centering
  \includegraphics[width=\linewidth]{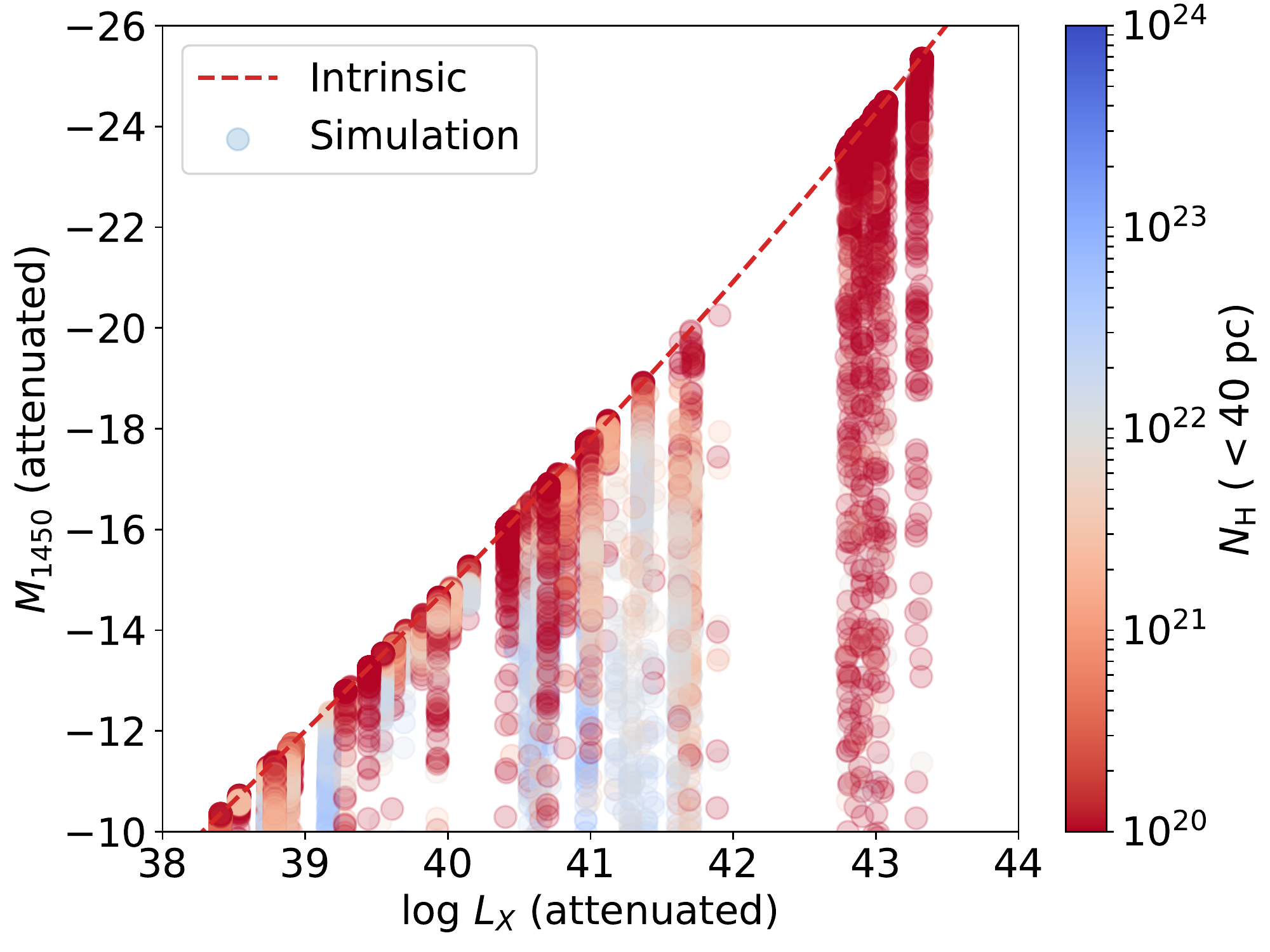}
  \caption{Mock UV magnitude at 1450 \AA{} versus the mock X-ray luminosity, colour-coded by \NH. The dashed red line indicates the relation between $M_{1450}$ and $L_{\rm X}$ assuming zero obscuration. High $L_{\rm X}$ corresponds on average to more massive BH in more evolved galaxies, therefore any obscuring material will be dust enriched and drastically reduce the observed UV.}
  \label{fig:LX_MUV}
\end{figure}

We can then compute an attenuated UV or X-ray luminosity. We illustrate this with Fig.~\ref{fig:LX_MUV}, which shows the UV magnitude as a function of X-ray luminosity, colour-coded by the local \NH. The red dashed line shows the intrinsic emission (i.e. the bolometric luminosity directly converted in X or UV). The deviations from this line correspond to differential attenuation between the X and the UV due to the presence of dust, and are more significant at higher luminosity. This is due to the fact that on average, the AGN in our simulation is brighter when the galaxy is more massive (and therefore more metal rich), so that the same \NH will correspond to a higher dust obscuration. However, even at low \NH and low luminosity, the UV luminosity can be significantly reduced.

\subsection{AGN duty-cycle}
\label{sec:agn-duty-cycle}

\begin{figure}
  \centering
  \includegraphics[width=\linewidth]{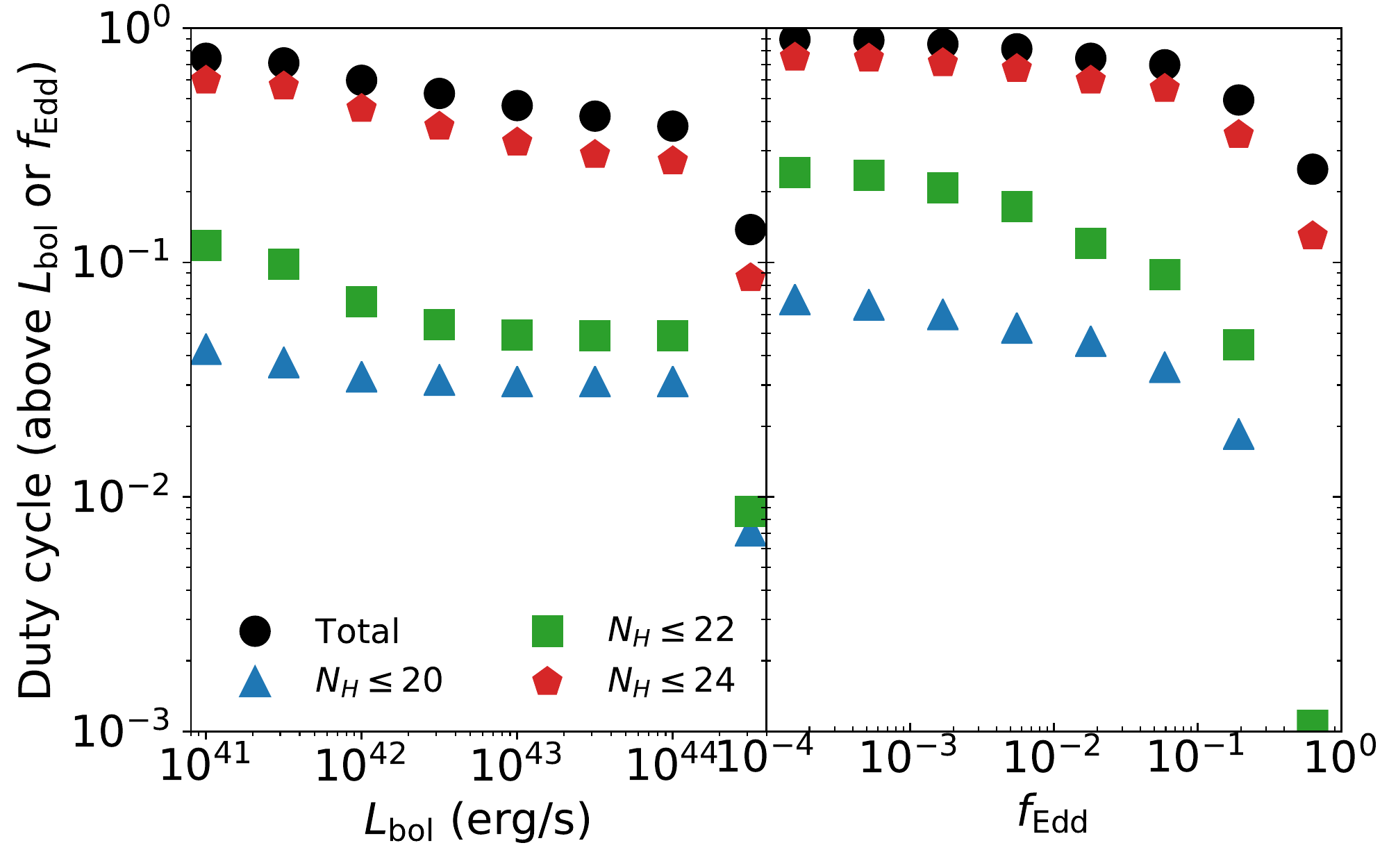}\\
  \includegraphics[width=\linewidth]{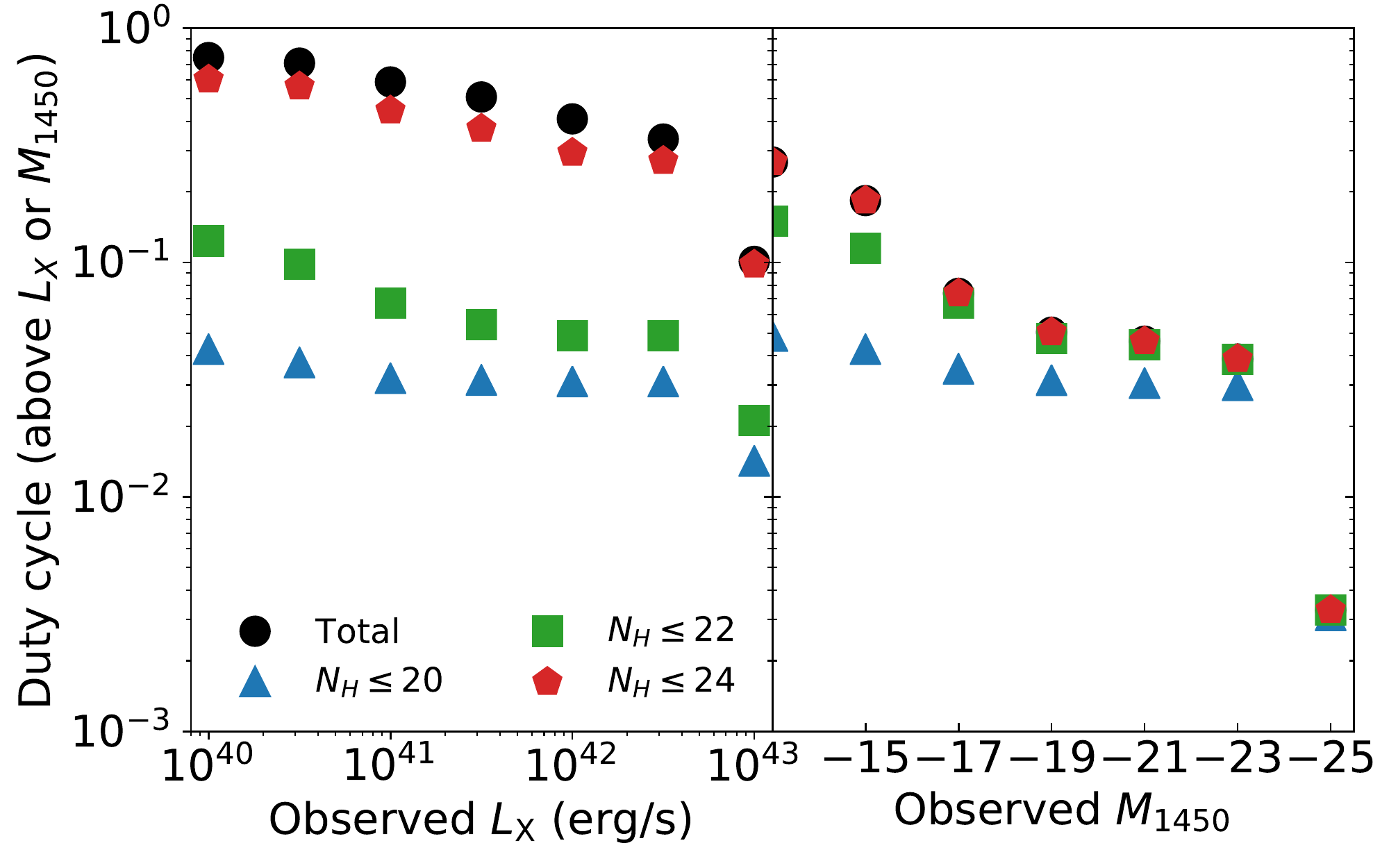}\\
  \caption{Intrinsic and observable duty-cycles for the BH in our simulation. For all panels, the black circles correspond to all the timesteps and all the directions around the BH, blue triangle select only fully unobscured BH (directions where $\log\NH \leq 20$), green squares are partially obscured BH ($\log\NH \leq 22$) and red pentagons correspond to all Compton-thin AGN ($\log\NH \leq 24$).
    \emph{Upper panel}: fraction of the time spent above a limiting $\Lbol$ (left) or \fedd (right). 
  \emph{Lower panel}: fraction of the time spent above a limiting $L_{\rm X}$ (left) or $M_{1450}$, taking into account obscuration by the gas in the galaxy and dust attenuation (right).}
  \label{fig:dutycycle}
\end{figure}

This brings us to distinguish between an ``intrinsic duty-cycle'', i.e., the fraction of time the BH is actively growing, and an ``observable duty-cycle'', i.e., the fraction of time that a growing BH can be observed as such. If the phases when the BH is most active are characterized by high obscuring column densities (see Figs.~\ref{fig:fracNH-distr} and~\ref{fig:Lbol-NH}), then the measurable growth time will be lower than the intrinsic growth time. The upper panel Fig.~\ref{fig:dutycycle} shows the fraction of time the BH is active, above a bolometric luminosity or Eddington ratio threshold, and compares the intrinsic duty-cycle to the one that would be inferred from observations. The lower panel shows the corresponding duty-cycle making use of the estimates of $L_{\rm X}$ and $M_{1450}$ discussed in the previous section.
\citet{2017MNRAS.465.1915R} show that the optical AGN luminosity function is in good agreement with the X-ray luminosity function of unobscured AGN ($\NH < 10^{21}\,\mbox{cm}^{-2}$), therefore we can use $\NH$ ranges to estimate duty-cycles in the UV/optical and X-rays. 

For this particular BH, it would be accreting above $\fedd > 0.2$ for around 50\% of the time, and $\sim 70$\% of this time (i.e. $\sim 35 \%$ of the total cosmic time) would be characterized by $\NH \leq 10^{24}\,\mbox{cm}^{-2}$. In other words, as expected, X-ray surveys would miss less than $\sim$ 30\% of the growth time.
Conversely, fully unobscured ($\NH< 10^{20}\,\mbox{cm}^{-2}$) growth above $\fedd > 0.2$, that we can associate as discussed above to optical visibility, represents less than 2-3\% of the total cosmic time. Requiring a growth above $\fedd > 0.5$ pushes this fraction far below 1\% of the cosmic time.
This is maybe even more noticeable on the lower panel of Fig.~\ref{fig:dutycycle}: while this specific AGN would be observed with an X-ray luminosity above $3\times 10^{42}\, \rm erg\,s^{-1}$ for approximately a third of the cosmic time, it would be fully unobscured for only 10\% of that time (around 3\% of the total time). This translates into the UV: the AGN only spends around 2-3\% of the time brighter than $M_{1450} \leq -19$, irrespective of the gas column density.

This explains why theoretical models which connect the halo mass function to the optical quasar luminosity function require very low duty-cycles, of order 0.01 \citep[e.g.,][]{2000ApJ...531...42H,2001ApJ...547...27H,2013MNRAS.428..421S,2018MNRAS.478.5564B} while timescale arguments require duty-cycles of order of unity to explain the masses of $z\sim6$ quasars \citep[e.g.,][]{2001ApJ...552..459H}.
Our results are in agreement with, and confirm, their assumptions.
Interestingly, the constraints on the quasar lifetimes derived by \citet{Chen2018} are extremely sensitive to the kinematics and distribution of cold gas in quasar hosts.

\section{Summary and conclusions}
\label{sec:ccl}
In this work, we have used a cosmological zoom-in simulation of a galaxy at $z \sim 6$, performed with the radiative hydrodynamics code \ramses, to investigate the link between AGN obscuration by gas on the scale of a few tens of parsecs and accretion onto BHs in high-$z$ galaxies. 

Our main results are the following:
\begin{itemize}
\item AGN feedback efficiently regulates the growth of the central BH by controlling the gas reservoir available for accretion (Figs.~\ref{fig:mstar-mbh} and \ref{fig:NH-fedd_EbEinj}).
\item Black holes accreting efficiently tend to be buried under a large column of gas more often than the ones accreting slowly (Fig.~\ref{fig:fracNH-distr}).
\item At high redshift, the gas in the galaxy and the circum-galactic medium contributes significantly to the total column density along the line of sight of an AGN (Figs.~\ref{fig:NHloc_NHgal} and \ref{fig:porosity}) 
\item Consistent with the findings of \citet{Vito2018}, the obscured fraction (Fig.~\ref{fig:fobsc_Vito18}) seems to be very high at high redshift and shows no clear trend with luminosity.
\item The UV emission from our high-$z$ AGN is more attenuated when the galaxy is more massive / metal rich (Fig.~\ref{fig:tauX_taudust}) and therefore at higher (observed) X-ray luminosity (Fig.~\ref{fig:LX_MUV}).
\item The fraction of time our BH is active and visible in the UV/optical is very low (of order 2\%), while its intrinsic growth duty-cycle is rather of order 50\%: the difference is due to the presence of highly obscuring gas in the galaxy (Fig.~\ref{fig:dutycycle}).
\end{itemize}

Despite the limited observability of an AGN in UV/optical based on its luminosity, the presence of an AGN in a galaxy can be revealed through emission line diagnostics \citep[e.g.][]{Baldwin1981, Juneau2011, Feltre2016}, a promising tool as shown by recent simulations \citep{Hirschmann2017}. The combination of deep X-ray surveys, as we confirm here, and emission line diagnostics, especially in the upcoming JWST era, will contribute to derive a more complete census of BH growth in high redshift galaxies.

\section*{Acknowledgements}

MT and MV acknowledge funding from the European Research Council under the European Community's Seventh Framework Programme (FP7/2007-2013 Grant Agreement no. 614199, project `BLACK').
This work has made use of the Horizon Cluster hosted by Institut d'Astrophysique de Paris; we thank St{\'e}phane Rouberol for running smoothly this cluster for us.
This work was granted access to the HPC resources of CINES under the allocation A0040406955 made by GENCI.
This work has made extensive use of the \texttt{yt}\footnote{\label{fn:yt}\url{https://yt-project.org/}} analysis package \citep{Turk2011}, as well as the \textsc{Matplotlib} \citep{Hunter2007}, \textsc{HealPy}, \textsc{Numpy/Scipy} \citep{Jones2001} and \textsc{IPython} \citep{Perez2007} packages.




\bibliographystyle{mnras}
\bibliography{bhgrowth} 



\appendix


\bsp	
\label{lastpage}
\end{document}